\documentclass[12pt,english,twoside]{article}

\usepackage{AST}
\usepackage{amsfonts}
\usepackage{bookmark}
\usepackage{amssymb}
\usepackage{amsmath}
\usepackage{mathrsfs}
\usepackage{graphicx}
\usepackage{mathtools}
\usepackage{hyperref}
\usepackage{lipsum}
\usepackage[labelfont=bf]{caption}
\usepackage[utf8]{inputenc}
\hypersetup{colorlinks=true,
    linkcolor=blue,
    filecolor=blue,
    urlcolor=blue,
    citecolor=blue
    }

\usepackage{float}
\usepackage[longnamesfirst,sort]{natbib}
\usepackage{qrcode}
\definecolor{darkblue}{rgb}{0,0,0.70}

\bibpunct{(}{)}{,}{a}{,}{,}
\usepackage[normalem]{ulem}

\newcommand{\SR}{{\rm I\kern-.21em R}}

\newcommand{\eenn}{\end{eqnarray}}

\def\ds{\displaystyle}

\def\player(#1,#2){\langle#1,#2\rangle}

\newcommand\blfootnote[1]{%
  \begingroup
  \renewcommand\thefootnote{}\footnote{#1}%
  \addtocounter{footnote}{-1}%
  \endgroup
}
\newcommand{\qqq}{\theta}
\newcommand{\eee}{\varepsilon}
\newcommand{\zzz}{\zeta}
\renewcommand{\lll}{\lambda}
\newcommand{\ddd}{\delta}

\renewcommand{\ggg}{\gamma}

\input tcilatex

\begin{document}

\title{
 \fontsize{32}{24}\selectfont{When is Trust Robust?$^{*}$}}
\author{\href{http://www.anderlini.net}{Luca Anderlini}\\
\begin{tabular}{c}\textit{\href{http://www.georgetown.edu}{Georgetown University} and}\\
\textit{\href{http://www.unina.it/en_GB/home}{University of Naples Federico II}}
\end{tabular}
\and \href{https://economics.yale.edu/people/larry-samuelson}{Larry Samuelson}\\
\textit{\href{https://www.yale.edu}{Yale University}}
\\
 \vspace{0.45em}
 \and \href{https://www.eief.it/eief/index.php/research/90-about-us/176-daniele-terlizzese}{\hspace{0.25em}Daniele Terlizzese}\\
\textit{\href{https://www.eief.it/eief/}{EIEF}}\vspace{0.4em}}

\date{\vspace{4em}{November 16, 2024}}

\maketitle
\keywords{Trust, Robustness, Fragility, Assimilation, Disruption\hspace{-0.3em}
\vspace{0.2em}
 } 

\jel{\href{http://www.aeaweb.org/econlit/jelCodes.php?view=econlit}{C72},
\href{http://www.aeaweb.org/econlit/jelCodes.php?view=econlit}{C79},
\href{http://www.aeaweb.org/econlit/jelCodes.php?view=econlit}{D02},
\href{http://www.aeaweb.org/econlit/jelCodes.php?view=econlit}{D80}
$\,$\hspace{-0.8em}
\vspace{0.2em}
 } 

\simplecorrespondent{\href{https://economics.yale.edu/people/larry-samuelson}{Larry Samuelson} $\;$ --- $\;$
\href{mailto:larry.samuelson@yale.edu}{{\tt Larry.Samuelson@yale.edu}}}

\runninghead{\textsc{When is Trust Robust?}}

\runningauthor{\textsc{Anderlini, Samuelson and Terlizzese}} 

\definecolor{darkblue}{rgb}{0,0,0.35} 
\vspace{3em}
%
%
\begin{abstract} \hspace{-1.1em}We examine an economy in which interactions are more productive if agents can trust others to refrain from cheating.  Some agents are scoundrels, who cheat at every opportunity, while others cheat only if the cost of cheating, a decreasing function of the proportion of cheaters, is sufficiently low.  The economy exhibits multiple equilibria.  As the proportion of scoundrels in the economy declines, the high-trust equilibrium can be disrupted by arbitrarily small perturbations or by arbitrarily small infusions of low-trust agents, while the low-trust equilibrium becomes impervious to perturbations and infusions of high-trust agents.  Scoundrels may thus have the effect of making trust more robust.%
\blfootnote{$^{*}$We thank the editor and three referees for helpful comments and suggestions.  Part of this research was done while Luca Anderlini and Larry Samuelson were visiting EIEF in Rome.  They are both grateful to EIEF for its hospitality.} 
\vspace{2em}
\end{abstract}


\section{Introduction}

Trust is important.   Trust can also be fragile---it can be laborious to build, easy to destroy, and difficult to rebuild.  As Mr. Darcy explains to Elisabeth Bennett in Jane Austen's {\em Pride and Prejudice}, ``My good opinion once lost is lost forever.''  This paper examines  conditions under which trust is fragile and conditions under which it is  robust.  

We study  an economy in which interactions are more productive if agents on one side of an interaction (proposers) can trust those on the other side (receivers) to refrain from cheating, and agents on the other side indeed do not cheat.   A typical receiver is ``responsive,'' meaning that the agent cheats if and only if the combination of a private and social cost of cheating is sufficiently low.  However, some receivers are  ``scoundrels'', who always cheat. We hereafter refer to the two varieties of receivers simply as  ``responsives'' and ``scoundrels.''   

The social cost of cheating is proportional to the probability that the cheater is either a scoundrel or a responsive who has violated the prevailing social norm.  The social norm is an equilibrium phenomenon specifying when it is acceptable to cheat, while the social cost represents the opprobrium heaped on a person who cheats when doing so is unacceptable.  The probability that a cheater is a scoundrel or a responsive violating the social norm is given by Bayes’ rule.  Bayes’ rule then immediately implies that the social cost of cheating is a convex and decreasing function of the equilibrium prevalence of cheating by responsives.

Because the social cost of cheating depends on the prevalence of cheating, multiple equilibria can arise.  If the fraction of scoundrels is sufficiently large,  there is a unique equilibrium in which no responsives cheat and trust is relatively high. If the fraction of scoundrels is smaller than a certain threshold, a high-cheating, low-trust (or ``bad'') equilibrium and a low-cheating, high-trust (or ``good'') equilibrium coexist (along with an unstable equilibrium exhibiting intermediate levels of cheating and trust).  

We assess robustness in two ways.  First, we introduce a belief-based best response dynamic under which the good and the bad equilibria are locally asymptotically stable, surrounded by basins of attraction that depend on the fraction of scoundrels.  The smaller is the fraction of scoundrels, the smaller is the  increase in the common perception of cheating required to catapult the good equilibrium out of its basin of attraction.  The good  equilibrium thus exhibits less cheating when there are fewer scoundrels, but sits more precariously within a smaller basin of attraction.  In contrast, decreasing the fraction of scoundrels {\em increases} the prevalence of cheating in the bad equilibrium and expands its basin of attraction.

Next, we examine the implications of introducing into an economy, characterized by either the high-trust or low-trust equilibrium, a small mass of agents characterized by the (quite different) beliefs and behavior characteristic of the other equilibrium.   If the fraction of scoundrels is sufficiently small, then an arbitrarily small infusion of agents accustomed to the low-trust equilibrium can disrupt the high-trust equilibrium, while a large infusion of agents accustomed to the high-trust equilibrium is required to disrupt the low-trust equilibrium. 

The asymmetry in size of the invasions required to disrupt the good and bad equilibrium is not simply the flip side of the fact that, as the fraction of scoundrels shrinks, the basin of attraction of the good equilibrium shrinks. The asymmetry holds even when the unstable equilibrium ---the boundary between the basins of attraction of the good and bad equilibria ---is kept halfway between the two equilibria.  The fraction of agents accustomed to the low-trust equilibrium required to disrupt the high-trust equilibrium is then considerably smaller than the fraction of those accustomed to the high-trust equilibrium required to disrupt the low-trust one. 

There is thus a sense in which scoundrels serve a useful purpose. A society in which  scoundrels are rare  is one in which a good social norm can be easily disrupted, while a bad social norm is more resilient. The best outcome is to have few scoundrels and coordinate on the good equilibrium, but this is fragile and risky. Tolerating some scoundrels may be a price worth paying for rendering the good equilibrium more robust.

Section \ref{model} presents the model, places our contribution in the  literature (most effectively done after seeing the basics of he model), and derives the equilibria of the model.  Section \ref{vitamin} characterizes the stability of the various equilibria and explains how this depends on the proportion of scoundrels.  Section \ref{robot} examines the robustness of the the good and bad equilibria to infusions of agents accustomed in each case to the other equilibrium.   Section \ref{discuss} interprets and discusses the results.  Proofs are gathered in an Appendix.  Any item with a number prefixed by ``A'' is to be found in the Appendix.


\section{The Model}\label{model}

\subsection{The Game}

The game is adapted from \cite{AandT2017}.  We view the game as capturing the spirit of the trust game of \cite{BDM95}, with the minimum modification required to ensure the equilibria can exhibit a positive level of trust.

In each period,  the members of a continuum of agents are matched into pairs to play a game.   Each time they are drawn to play the game, each agent is equiprobably assigned to be either a proposer or a receiver.  The proposer first chooses a quantity $x\in \mathbb R_+$.  The receiver then chooses either to cheat or not cheat.  If the receiver does {\em not} cheat, then proposer and receiver each receive a payoff of $x$.  If the receiver cheats, then the proposer receives 0.

Proportion $q$ of the receivers are \textit{scoundrels}, who cheat at every opportunity, and whose payoffs we accordingly need not specify. The \textit{responsives} are a proportion $1-q$ of the receivers. When a responsive cheats, she receives $2x$ minus the cost of cheating.  The fraction of scoundrels is known, but scoundrels are not distinguishable from responsives. 

We can interpret $x$ as a proposed scale at which to operate a joint project.  As the scale increases, so do the payoffs of both agents if they indeed share the proceedings, but so does the payoff to the receiver from cheating and thereby appropriating the entire payoff (minus the cost of cheating).

The cost of cheating is the sum of a private cost and a social cost that reflects a social norm.  

\subsection{The Private Cost of Cheating}\label{pcost}

In each interaction, a responsive has a private cost $z$ independently drawn from  the uniform distribution on $[0,1]$.%
\footnote{The uniform distribution simplifies the exposition and various calculations, but the qualitative results do not depend on this assumption.}
Our interpretation is that responsives have an intrinsic aversion to cheating, manifested in feelings of guilt and self-censure, that are attenuated when the need to cheat or the benefits from cheating are high.  A low value of $z$ thus reflects  a high need to cheat or high benefit from cheating, and hence a low private cost---receivers are willing to forgive themselves for cheating when the need or payoff is high.  A receiver's value of $z$ does not depend on the offer $x$ they receive---the offer $x$ affects the reward for cheating, but $z$ captures an intrinsic characteristic that is independent of $x$.  

Suppose, for example, that cheating takes the form of cutting ahead of  others in traffic.  We might think of a low $z$ as identifying a person who is on the way to the hospital while experiencing chest pain, and so has an urgent need for haste.  A medium value of $z$ might identify a person who is late for work, and so has a moderate need for haste.  A high value of $z$ is a person not pressed for time.     If instead cheating takes the form of jumping a queue, then a low value of $z$ is someone for whom an emergency renders immediate service imperative, while a high value of $z$ is someone not pressed for time.  Alternatively, suppose cheating takes the form of failing to make the honor-system payment at the workplace coffee machine.  A low $z$ may represent someone who has forgotten their wallet that day, needs a pick-me-up before an important meeting, and plans to repay the next day.%
\footnote{We can see counterparts of the offer $x$ in these scenarios.  In traffic, it takes the form of a willingness to yield to others.  In queuing, it is reflected in the willingness to form queues in the first place, rather than mob head of the line.  In the office, one sees it in the extent to which concessions are provided on the honor system.}

Scoundrels have no intrinsic aversion to cheating.  Whether in the midst of an emergency or at leisure, they feel no private cost.  Scoundrels are similarly oblivious of the social norm and impervious to the attendant social cost of cheating, and hence cheat at every opportunity.

A social norm determines when responsives find it acceptable to cheat, in the  form of a value $\zeta$, along with the view that it is socially acceptable for responsives with values $z<\zeta$ to cheat, and socially reprehensible for responsives with values $z>\zeta$ to cheat.  The value of $\zeta$ is an equilibrium phenomenon, reflecting a social consensus on what constitutes acceptable behavior.  Returning to our traffic example, some societies have coordinated on a low value of $\zeta$, and pedestrians can cross the street with impunity, confident that only in extreme cases (i.e.,  very low values of $z$) do drivers become so aggressive as to ignore pedestrians. Others have settled on a higher value of $\zeta$ and motorists routinely ignore pedestrian crossings. In the queuing example, in some societies $\zeta$ is low and queues are common and commonly respected.  In others, $\zeta$ is high and queues are routinely flaunted, to the extent that ``lines don't grow longer, they only grow thicker''.  

\subsection{The Social Cost of Cheating}\label{scost}

The social cost of cheating takes the form of public disapproval,  ostracism, or other forms of sanction.  This cost shows up in the disapproving looks, the clucking of tongues and muttered comments, the outright chastisement and sometimes worse, that follows the violation of a social norm.   If receivers' types were  observable, the social punishment would be directed only at responsives who violate the social norm, by cheating while having values $z>\zeta$, and at scoundrels.  However, types are not observable.  Instead, punishers form a belief about the type of a cheating receiver, and the severity of the punishment is proportional to the probability assigned to the receiver being either a responsive with $z>\zeta$ or a scoundrel. 

This estimation process is simplified by the observation that in equilibrium, responsives for whom $z>\zeta$ will never cheat.  This is an implication of the fact that the social norm $\zeta$ is itself an equilibrium phenomenon.  If responsives with values $z>\zeta$ found it optimal to cheat, then $\zeta$ would not persist as the social norm. Instead, the social norm would drift upward to match the prevailing behavior. The equilibrium condition for the social norm is that induced behavior indeed conforms to the norm. The severity of punishment is thus proportional to the probability that a cheating receiver is a scoundrel.   

The social cost of punishment depends on the social norm.  The traffic menace is sanctioned more severely when the equilibrium is such that the only responsives who cheat are the few imminently expectant mothers headed for the hospital, and hence cheaters are likely to be  scoundrels who routinely flaunt traffic conventions. To make this connection precise, let $s$ be the proportion of responsives who cheat.  The  posterior probability that someone observed cheating is a scoundrel is
\begin{eqnarray}\nonumber
\frac{q}{q+(1-q)s}.
\end{eqnarray}
We then take the social cost of cheating to be proportional to this probability, or
\begin{eqnarray}\label{brahms}
f(s)= \qqq \, \frac{q}{q+(1-q)s},
\end{eqnarray}
where $\theta>0$ is a parameter that allows us to tune the relative importance of the idiosyncratic and social components of the cost of cheating. The total cost of cheating for a responsive of type $z$, denoted by $c(z,s)$ is then 
\begin{eqnarray}\label{eqn: total cost of cheating defined}
c(z,s)=z + f(s) = z+\qqq \, \frac{q}{q+(1-q)s}.
\end{eqnarray}

\subsection{Relation to the Literature}\label{subsection: related literature}

Our point of departure is the belief that trust is important.  \cite{arrow1974limits} argued that even the simplest of economic transactions calls for a foundation of trust.%
\footnote{Arrow's argument was illustrated by the classic empirical study by \cite{Banfield:58}, documenting the effects of the lack of trust on a small community in southern Italy. The ``amoral familism'' that stems from the lack of trust has calamitous effects on that ``backwards'' society. Italy is also the object of \cite{Putnam:93}'s investigation of the role of different levels of social capital and their effects on democracy.} 
\cite{Fukuyama:95} provided a famously optimistic view of the effects of trust on large firms and overall growth.%
\footnote{\cite{Fukuyama:95}'s optimism found some notable skeptics, including \cite{Solow:95}.}
\cite{levitsky2019democracies} argue that  democracies require two ingredients to function 
effectively, namely that competing parties accept one another as legitimate rivals and that they trust one another to exercise restraint in exploiting their institutional advantages.   \cite{bowles2016} argues that a society can function well only if people can trust one another to follow social norms.  A large literature, catalyzed by \cite{Putnam:00}, with \cite{Jackson:20} providing a recent point of entry, explores the link between social capital, defined in various ways but routinely including some component of trust, and economic development.   For an early survey, see \cite{Sobel:02}.

Trust can also be fragile.  The folk wisdom that trust can be laborious to build, easy to destroy, and difficult to rebuild is backed up by research in psychology.  See  \cite{slovic1993perceived,slovic1999trust} for influential early studies and \cite{doyle2023fragile} for a recent contribution.   

Our incorporation of a private and social cost of cheating places our paper in a literature that relies on a specification of social preferences---some form of a cost of cheating, altruism, reciprocity, inequality aversion, a concern for esteem or the good opinion of others, and so on---to generate trust.%
\footnote{Starting with the theoretical work of \cite{Kandori:92} and continuing with theoretical and experimental papers such as  \cite{Xie-Lee:12}, \cite{Duffy-Xie-Lee:13} and \cite{DalBo-Frechette:18} among others, a literature has examined an alternative approach in which trust or social norms are sustained by repeated interactions.  \cite{Mailath-Samuelson:06} provide an introduction to the literature.  } 
Our work is especially connected to the subset of this literature in which the social preferences include some concern for how an agent is perceived by others.  In a similar vein, \citet{Benabou-Tirole:06} examine a model in which agents are motivated by a combination of altruism, extrinsic incentives, and a concern for esteem.  \citet{Tadelis:11} studies a trust game in which agents are motivated partly by a concern that others perceived them as trustworthy.  \citet{Andreoni-Bernheim:09} examine a dictator game in which proposers have a concern for fairness and for being perceived as fair.  \citet{Ellingsen-Johannesson:08} examine the effects of the desire for ``esteem'' in a principal-agent setting.  Of course, the voluminous signaling literature is concerned with settings in which an agent cares about how she is perceived by others.  Our social cost of cheating, proportional to the posterior belief about the agent's type, is a standard objective for senders in signaling models.  We differ from much of the signaling literature in that the updating leading to this belief is an endogenous function of the actions of others besides the sender, and by the fact that the sender signals in order to separate from an undesirable type, rather than pool with a desirable type.  \cite{Mailath-Samuelson:01} provide a discussion of signaling to separate.

The social preferences approach to trust was energized by the theoretical and 
experimental work of \cite{BDM95}, in a game whose unique subgame-perfect equilibrium exhibits neither trust nor trustworthiness.%
\footnote{\cite{johnson2011trust} and \cite{naef2009can} examine the subsequent literature on trust games.}  
We retain the spirit of their game, but follow the lead of  \cite{AandT2017} in modifying the game by introducing the cost of cheating described in Sections \ref{pcost} and \ref{scost}.   The key effect of this modification is that positive levels of trust now emerge in equilibrium, allowing us to study the relative fragility or robustness of equilibrium trust.

As explained in Sections \ref{pcost} and \ref{scost}, we interpret the equilibrium trust that arises in our model as a social norm.  For introductions to the extensive literature on social norms and social preferences  see \cite{burke2011social} and \cite{postlewaite2011social}.   
We have constructed our model so that the norm behavior takes a particularly simple form.  
The absence of noise in our model leads to  equilibria in which no deviations from the norm on the part of the responsives are observed.  This simplifies the robustness examination and the exposition.  Once the model is enriched so that norms are broken in equilibrium the issue of how they are (or should be) enforced becomes more delicate. \cite{Acemoglu-Jackson:17} analyze a model in which there is a complex interplay between norms and laws intended to enforce them. Among other things, they show that laws that are too ``tight'' relative to social norms may backfire, in the sense of being less effective than ``gradual'' enforcement.    \cite{bowles2016} expresses similar sentiments.

Throughout  this discussion of the literature and throughout the paper, we maintain the spirit of the  Arrovian view of norms as a way to achieve otherwise elusive efficiency gains.  This view has attracted a critical minority view.  \cite{Elster:89} argues  that ``many social norms do not benefit anyone.''  Prime examples are exclusionary social rules and/or rules that forbid certain types of behavior.  By contrast, a higher level of trust is beneficial to all in our economy.

\subsection{Preliminaries}\label{subsection: preliminaries}

Using (\ref{eqn: total cost of cheating defined}), the payoff of a responsive who cheats is given by
\[
2x-c(z,s).
\]
If $s$ is small, then a cheater is likely to be a scoundrel, and cheating will be punished heavily.  If $s$ is large, then it is relatively unlikely that a cheater is a scoundrel,
and cheating will be only lightly punished. 

\begin{figure}[H]
	\center\includegraphics{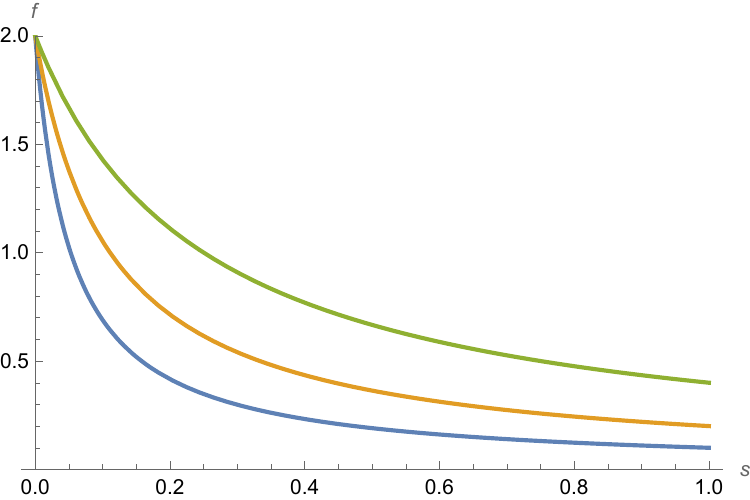}
	\caption{Illustration of the social cost of cheating $f(s)$ as a function of the proportion $s$ of 
		responsives who cheat, for $\theta = 2$
		and the proportion of scoundrels (top to bottom) $q = 0.2, 0.1, 0.05$.\label{dungaree}}
\end{figure}

We note that  the function $f:[0,1]\rightarrow \mathbb R_+$ is a decreasing, convex function with $f(0)=\qqq$ and $f(1) = \qqq q$.
Therefore the social component of the cost of cheating is maximal when no responsive cheats, falls quickly as soon as a few of them cheat,
and keeps falling, but at a decreasing rate, as more and more responsives cheat.  The fewer  the scoundrels, i.e. the smaller is $q$, the greater is the convexity of $f$. In particular, the steeper is $f$ near 0.\footnote{The posterior probability that a cheater is a scoundrel is 1 when $s=0$. In order for it to fall from 1 to $\frac{1}{2}$, $s$ must increases  from 0 to $\frac{q}{1-q}$. Hence, the increase in $s$ that generates such a decrease becomes smaller as $q$ decreases.}
Figure \ref{dungaree} illustrates.

As $q$ approaches zero, the function $f$ converges (but not uniformly) to
\begin{eqnarray}\nonumber
\begin{array}{rcl}
	f(0) &=& \theta\\
	f(s) &=&0 \quad{\rm for} \quad s\; >\; 0.
\end{array}	
\end{eqnarray}

 \subsection{Equilibrium}

A responsive takes the proportion $s$ of responsives who cheat as given, and when facing  an offer $x$, will cheat if her cost of
cheating $z$ falls short of a cutoff $\zzz(x,s)$ and will not cheat if  $z\geq \zzz(x,s)$.  The cutoff $\zzz(x,s)$ equalizes the payoffs of cheating and not cheating, and
hence when interior solves
\begin{eqnarray}\nonumber
2x-[\zeta(x,s)+f(s))] = x.
\end{eqnarray}
In general, we have
\begin{eqnarray}\label{senate}
	\zzz(x,s) = \max\{0, x-f(s)\}.
\end{eqnarray}
The maximum reflects the possibility of a corner solution in which the responsive does not cheat even if $z=0$.  In principle we could also have a corner solution in which the responsives cheat even if $z=1$.  In the next two paragraphs we will see that this does not arise.

A proposer takes the proportion $s$ of responsives who cheat as given and chooses a value $x$ to maximize the payoff
\begin{eqnarray}\nonumber
(1-((1-q)\zzz(x,s)+q))x,
\end{eqnarray}
where $1-((1-q)\zzz(x,s)+q)$ is the (overall) probability that the current receiver does {\em not} cheat.  Using  \eqref{senate}, we can write the maximization problem of a proposer as
\begin{eqnarray}\nonumber
\max_{x\geq 0} (1- ((1-q)\max\{0, x-f(s)\}+q))x.
\end{eqnarray}

The proposer will never set $x\ge f(s)+1$. Doing so would induce all responsives to cheat and hence would yield a payoff of 0, while the proposer can ensure a positive payoff by setting  $x<f(s)+1$.  Equivalently, we will never have a corner solution in which all responsives cheat.
The proposer will similarly never set $x<f(s)$, since doing so would ensure that no responsives would cheat, {\em and} that the proposer could increase the offer without inducing additional cheating.

We can thus restrict attention to offers $x\in [f(s),f(s)+1)$.  The proposer's objective is then to solve
\begin{eqnarray}\nonumber
\max_{x\in [f(s),f(s)+1)} (1- (1-q)(x-f(s))-q)x.
\end{eqnarray}
The first-order condition if $x>f(s)$ is
\begin{eqnarray}\label{streetlight}
1+f(s)-2x=0 \;\; \Longleftrightarrow \; x=\ds\frac12+\ds\frac12 f(s).
\end{eqnarray}
This is the relevant solution as long as $x>f(s)$, i.e. as long as $\frac12+\frac12 f(s)>f(s)$,
or, equivalently, as long as $f(s)<1$. Let $s^*$ be the solution to  $f(s^*)=1$.   Using (\ref{brahms}),  we can solve $f(s^*)=1$ to obtain
\begin{eqnarray}\label{s star defined}
	s^*=\frac{q (\theta -1)}{1-q}.
\end{eqnarray}
When  $\theta\ge 1$, we can interpret $s^*\ge 0$ as a proportion of responsives who cheat.  If $s>s^*$, the proposer will then choose an interior solution (satisfying \eqref{streetlight}) in which some responsive cheat.  If $s<s^*$, cheating is sufficiently costly that the proposer finds it optimal to deter all responsive cheating by choosing the highest value of $x$ consistent with no such cheating, namely $x=f(s)$.  If $\theta<1$,  the solution \eqref{s star defined} exists but is negative, thus defying interpretation as a proportion of cheaters.  In this case \eqref{s star defined} is not relevant and the proposer always chooses an interior solution.

We thus have
\begin{eqnarray}\nonumber
x =
\left\{
\begin{array}{lll}
	\ds\frac12 \; + \; \ds\frac12f(s)	&&s\ge s^*\\
	& &\\
	f(s)&&s\le s^*.
\end{array}	
\right.
\end{eqnarray}

The equilibrium condition is that the proportion $s$ of cheating by responsives must induce a proposer offer $x$ that in turn causes the cutoff $\zzz(x,s)$ to match $s$.  We thus have three conditions which jointly determine the equilibrium values of $s$, $\zzz$ and $x$:
\begin{eqnarray}
	s&=&\zzz(x,s) \label{eric}
\end{eqnarray}
\begin{eqnarray}\label{jack}
	\zzz(x,s) &=& \max\{0, x-f(s)\}
\end{eqnarray}	
\begin{eqnarray}\label{ginger}
	x &=&
	\left\{
	\begin{array}{lll}
		\ds\frac12+\ds\frac12 \, f(s)	& &s\ge s^*\\
		&&\\		
		f(s)&&s\le s^*
	\end{array}	
\right.
\end{eqnarray}	
The final condition (\ref{ginger}) can be rewritten as
\begin{eqnarray}\nonumber
x=\max \left\{f(s),\frac12+\frac12f(s)\right\}.
\end{eqnarray}

\subsection{Equilibrium Cheating}

Our first result is that if the social cost of cheating is sufficiently low, then there is a unique equilibrium, which exhibits some cheating.  The proof, contained in Section \ref{mahler}, is a straightforward calculation. The left panel of Figure \ref{park} below illustrates this case.

\begin{proposition}\label{cantata}
If $\qqq<1$, there exists a unique equilibrium.  In equilibrium, some responsives cheat.	
	\end{proposition}

We are interested in the case of multiple equilibria.  We accordingly assume throughout the following, without subsequent mention, that the social component of the cost of cheating is sufficiently important:
\begin{ass}
$\theta >1$
\end{ass}
\noindent In this case, one corner-solution equilibrium configuration is
\begin{eqnarray}\nonumber
s=\zzz=0,~~~~x=f(0)=\theta.
\end{eqnarray}
This is a high-trust, no cheating equilibrium, featuring a relatively large offer $x$ and no cheating on the part of responsives.  Given the assumption that $\theta >1$, this equilibrium always exists.  We refer to this as the good equilibrium, and denote the proportion of responsives who cheat in this equilibrium by $s_g=0$.

 If the social cost of cheating $f(s)$ decreases sufficiently rapidly in $s$, then we have two additional, interior solutions.  Each of these must satisfy $s\ge s^*$, and hence must satisfy the interior versions of \eqref{eric}--\eqref{ginger}, or $\zzz(x,s)=x-f(s)$ and $x=\frac12+\frac12f(s)$.  We can reduce \eqref{eric}--\eqref{ginger} to a single equation in $s$, given by
\begin{eqnarray}\nonumber
\frac12+\frac12f(s)=s+f(s),
\end{eqnarray}
which in turn can be rearranged to read
\begin{eqnarray}\label{bandit}
f(s) = 1-2s.
\end{eqnarray}
Given the specification of $f(s)$ as in (\ref{brahms}) this is a quadratic equation, whose solutions are
%
\begin{eqnarray}\label{s_b}
	s_b=\frac{1-3q+\sqrt{(q+1)^2-8\theta q(1-q)}}{4(1-q)}
\end{eqnarray}
\noindent and
\begin{eqnarray}\label{s_u}
	s_u=\frac{1-3q-\sqrt{(q+1)^2-8\theta q(1-q)}}{4(1-q)}.
\end{eqnarray}
We thus have a low-trust equilibrium in which proportion $s_b$ of responsives cheat, and an intermediate equilibrium in which proportion $s_u$ of responsives cheat.  We refer to these as the bad equilibrium and the unstable (for reasons made clear in Section \ref{vitamin}) equilibrium, respectively.

The bad and unstable equilibria  exist if the expression under the square root in (\ref{s_b})---(\ref{s_u}) is positive.  This is true if there are not too many scoundrels, with the upper threshold on the fraction of scoundrels given by
\begin{eqnarray}\label{mozart}
	\hat{q}(\qqq)=\frac{4 \theta-1-4\sqrt{\theta(\theta-1)}}{1+8 \theta}.
\end{eqnarray}
If $q<\hat{q}(\qqq)$, we have $0=s_g<s_u<s_b$.  If $q > \hat{q}(\qqq)$, then no cheating is the only solution.%
\footnote{The discriminant would also be positive  if $q$ were larger than the larger solution of the quadratic. In this case, however, both $s_b$ and $s_u$ would be negative. Therefore, only the smaller solution of the quadratic is relevant.  	For the boundary case of $q=\hat q$, the positive solutions $s_b$ and $s_u$ exist and coincide.}

\begin{figure}[H]
\begin{center}\includegraphics[scale=0.5]{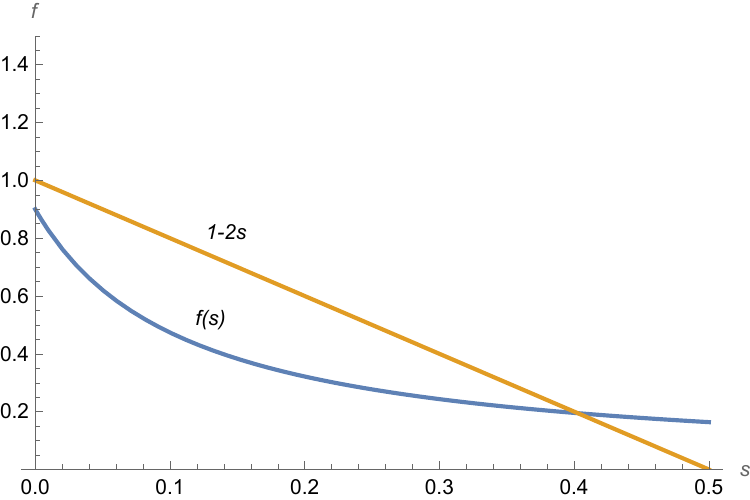}	
\includegraphics[scale=0.5]{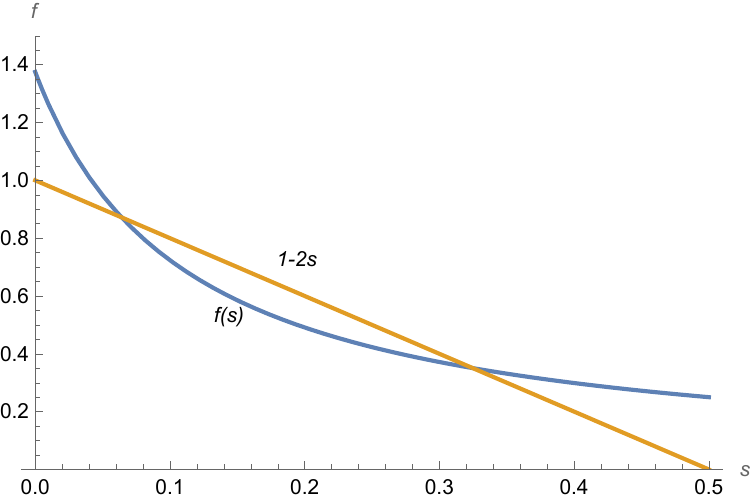}
\end{center}	
\caption{Illustration of equilibria.  In the left figure, $\qqq=0.9$ (the social cost of cheating is relatively low), and there is a single, interior equilibrium.  In the right picture, $\qqq=1.375$ and there are relatively few scoundrels ($q=0.1$ in both panels), giving rise to three equilibria.  The high-trust equilibrium $s=0$ corresponds to the intersection of $f(s)$ with the vertical axis, while the intermediate and low-trust equilibria are determined by the two interior intersections.  As the proportion of scoundrels increases, the function $f$ shifts upward, pushing the intermediate and low-trust equilibria closer together, until a point is reached at which $q= \hat q(\qqq)$ and these equilibria first coincide and then disappear, leaving only the high-trust  equilibrium. \label{park}}
\end{figure}

We summarize with the following proposition, illustrated in Figure \ref{park}:%
\footnote{\label{coffee}If the social cost of cheating function $f$ was concave instead of convex and $\theta$ was sufficiently large, then we would have $f(s)>1-2s$ for all $s$, and a single equilibrium on the vertical axis, in which no responsives cheat.  For yet smaller values of $\theta$, the functions $f(s)$ and $1-2s$ may exhibit two interior intersections, analogously to the right panel in Figure \ref{park}, but now with a stable, interior low-cheating equilibrium, an unstable, interior intermediate equilibrium, and a stable high-cheating equilibrium on the horizontal axis.  In between, we expect a configuration in which  $f(s)$ and $1-2s$ intersect once, analogously to the left panel in Figure \ref{park}, whose nature depends on the specification of the function $f$.  If $f$ cuts $1-2s$ from above, then there are three equilibria, including a stable equilibrium on the vertical axis in which no responsives cheat, an unstable interior equilibrium, and another stable, high-cheating equilibrium on the horizontal axis. If $f$ cuts $1-2x$ from below, then there is a unique, interior equilibrium.}

\begin{proposition}\label{the good the bad and the ugly}
	
[\ref{the good the bad and the ugly}.1]  The good equilibrium is the unique equilibrium  if $q>\hat q(\theta)$, where the function $\hat q(\qqq):[1,\infty)\rightarrow [0,1]$  
is decreasing, and
\begin{eqnarray*}
	\hat q(1) \;=\; \frac{1}{3} \qquad {\rm and} \qquad
	\lim_{\qqq\rightarrow \infty}\hat q (\qqq)\; = \; 0.
\end{eqnarray*}	

\hspace{6em}[\ref{the good the bad and the ugly}.2] 
If $q<\hat q(\qqq)$, then in addition to the high-trust, no cheating (good) equilibrium, there is a  low trust, high cheating (bad) equilibrium in which a proportion $s_b$ of responsives cheat, and an intermediate (unstable) equilibrium in which a proportion $s_u$ of responsives cheat.%
\footnote{{\em There are two boundary cases.
		When $q=\hat q(\qqq)$, there exist only two equilibria, a stable equilibrium $s_g=0$ and another equilibrium (intuitively, $s_u=s_b$) that
		is stable from above but not from below.  When $q=0$, there exist only two equilibria, a stable equilibrium $s_b$ and an another, unstable equilibrium
		(intuitively, $s_g=0=s_u$).}} 

\hspace{6em}[\ref{the good the bad and the ugly}.3]
The offers made by proposers are the highest and cheating is the lowest in the good equilibrium, while offers are the lowest and cheating the most prevalent
in the bad equilibrium.
\end{proposition}

\section{Local Asymptotic Stability}\label{vitamin}

We now investigate the resilience of the high trust equilibrium, in two steps.  The first, examined in this section, asks about the stability of the various equilibria under a dynamic process in which agents play best responses to beliefs that adapt toward realized behavior.  

We characterize the state of the economy by a  \emph{perceived } level of cheating, denoted by $s_P$, common to everyone in the economy.   We interpret the commonality of the perception $s_P$ as reflecting access to common sources of information concerning the prevalence of cheating.  The media regularly reports information on the prevalence of crime, violations of social norms often cause disruptions that others can observe, incidents of cheating give rise to word-of-mouth chains of information, and so on.  The perception $s_P$ may or may not be an equilibrium level of cheating.  
 
Given a perception $s_P$, proposers choose the value of $x$ that would maximize their payoff if $s_P$ was the prevailing proportion of responsives who cheat.   Hence, from \eqref{ginger}, proposers choose 
\[
x = \max\left\{f(s_P), \frac12+\frac12f(s_P)\right\},
\]
Receivers similarly take $s_P$ to be the prevailing proportion of responsives who cheat and react to the offer $x$ by choosing to cheat (or not) in order to maximize their payoff.  From \eqref{jack}, these decisions  give rise to a realized proportion of cheating among responsives $s$ that solves 
\[	s = \max\{0, x-f(s_P)\}.
\]
We can rearrange these two equalities  to obtain the realized proportion of cheating, given by
\begin{eqnarray*}
	s=
	\left\{
\begin{array}{ll}	
0~~~~~~~~~~~~~~~~~~~&s_P\le s^*\\
	&\\
\displaystyle\frac12-\frac12f(s_P)~~~~~&s_P\ge s^*~.
\end{array}
\right.
\end{eqnarray*}
Proposers and receivers thus both choose best responses to their perceptions.

The potentially erroneous perception $s_P$ moves toward the induced realization $s$.    When the media reports, observations, and informal reports reflect behavior that differs from the prevailing perception $s_P$, perceptions adjust to move close to the evidence.  Our results do not depend on details such as whether this adjustment happens instantaneously, or quickly, or sluggishly.  We only require that if the prevailing perception $s_P$ is at odds with society's aggregate experience, then there will be pressure pushing the perception toward the experience.  

The movement of the perception $s_P$ in the direction of the realized incidence of cheating $s$ suffices to ensure that the dynamic has three rest points, $s_g$, $s_u$ and $s_b$.  A rest point $s$ is {\em locally asymptotically stable} if there exists a neighborhood of $s$, referred to as its basin of attraction, with the property that from any initial condition in this neighborhood, the dynamics converge to the state $s$.    Local asymptotic stability ensures that the dynamic process will converge to a rest point if its initial condition is nearby, as well as that a population incurring a small shock away form a rest point will return to the rest point.

\begin{figure}[H]
	\begin{center}	
		$\,$\hspace{-2.5em}\includegraphics[scale=0.40]{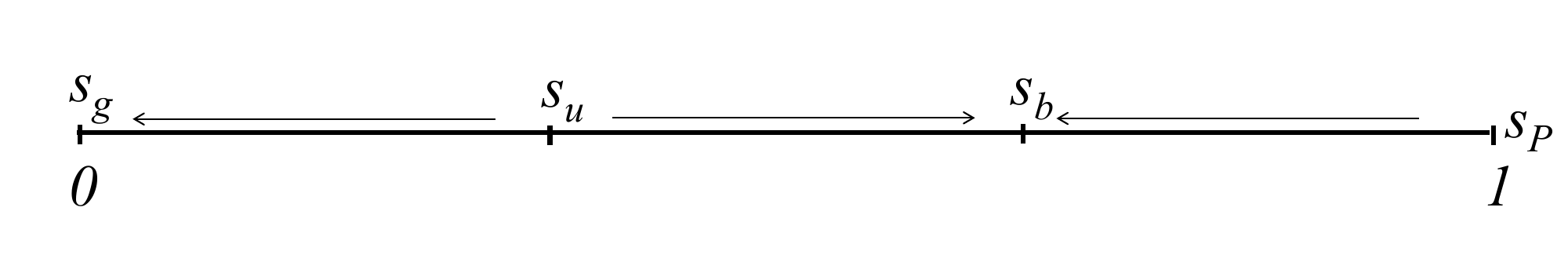}	
		\caption{Illustration of the adjustment dynamics and basins of attraction for the candidate equilibrium proportion $s_P$ of responsives who cheat.\label{orchestra}}	
	\end{center}
	\vspace{-1em}
\end{figure}

As long as $s_P\leq s^*$, the induced realization of $s$ is always 0, so $s_P$ falls towards 0.  If $s_P\in (s^*, s_u)$, we have that $s=\frac12-\frac12f(s_P)<s_P$, and therefore again $s_P$ falls towards 0.  If $s_P\in (s_u, s_b)$ the realized $s$ is larger than $s_P$, which therefore increases towards $s_b$.  Finally, if $s_P>s_b$, the realized $s$ is smaller than $s_P$, implying that $s_P$ falls back towards $s_b$.   Hence, the good equilibrium  $s_g$ and the bad equilibrium  $s_b$ are  locally asymptotically stable, while  the intermediate rest point $s_u$ is unstable.   The unstable equilibrium divides the interval $[0,1]$ of possible values of $s_P$ into the basin of attraction $[0,s_u)$ of the lower rest point $s_g$ and the basin of attraction $(s_u,1]$ of the upper rest point $s_b$. Figure \ref{orchestra} illustrates.

We can expect the initial conditions to be more likely to fall into the basin of attraction of $s_g$ (or, similarly, into the basin of attraction of $s_b$) the larger is this basin.  Similarly, an equilibrium is more likely to withstand shocks that push society away from it the larger is the distance from the equilibrium to the boundary separating its basin of attraction from that of the adjacent equilibrium.   We accordingly note that the good equilibrium sits distance $s_u-s_g$ from the relevant boundary of its basin of attraction $[0,s_u)$ and the bad equilibrium sits  $s_b-s_u$ away from the relevant boundary $s_u$ of its basin $(s_u,1]$.   The comparative statics in the following proposition, which is proved formally in Section \ref{app: proof of prop Sea},  are an immediate consequence of (\ref{s_b})--(\ref{mozart}).

\begin{proposition}\label{sea}
Assume that $q<\hat q(\qqq)$, so that all three equilibria exist.
As either $q$ or $\qqq$ increase, then $s_u$ increases and $s_b$ decreases, and hence
\[
s_u-s_g {\rm ~increases;~} ~~~ 1-s_u{~\rm and ~}s_b-s_u {\rm ~decrease~}.
\]
Conversely, as $q$ approaches $zero$, $s_u$ also approaches zero and hence  the basin of attraction of the good equilibrium $s_g$ becomes arbitrarily small and the basin of the bad equilibrium $s_b$ becomes approaches the entire interval."

\end{proposition}

Hence, when $q$ is small, there is relatively little cheating in the good equilibrium (since there are few scoundrels), but the good equilibrium is fragile, in the sense that it has a small basin of attraction, while cheating is relatively prevalent in the bad equilibrium.  As $q$ increases, so does the incidence of cheating in the good equilibrium, but the good equilibrium has a larger basin of attraction, while the incidence of cheating in the bad equilibrium decreases.  When $q$ hits $\hat q (\qqq)$, the unstable and bad equilibria coincide, and for larger values of $q$ only the good equilibrium remains, albeit with more scoundrels.  As $\qqq$ increases, the proportion of scoundrels needed to eradicate the unstable and bad equilibria decreases.

We now see two respects in which it can be ``good'' to have more scound\-rels.  First, the more scoundrels there are, the ``more likely'' is the good equilibrium to be the unique equilibrium (more precisely, the smaller is the value of the social-cost-of-cheating parameter $\qqq$ required to ensure the good equilibrium is unique).  Second, when multiple equilibria exist, the good equilibrium is ``more likely'' the more scoundrels there are (more precisely, the larger is the basin of attraction of
the good equilibrium).

Of course, scoundrels come at a cost---society has to put up with their cheating.  The most fortunate society is one that contains few scoundrels, but manages  to  coordinate on and preserve the good equilibrium, despite its fragility.  A less fortunate society is that which still has few scoundrels, but is trapped at the bad equilibrium.   

\begin{figure}[H]
\vspace{-0.5em}
\begin{center}\includegraphics[scale=0.16]{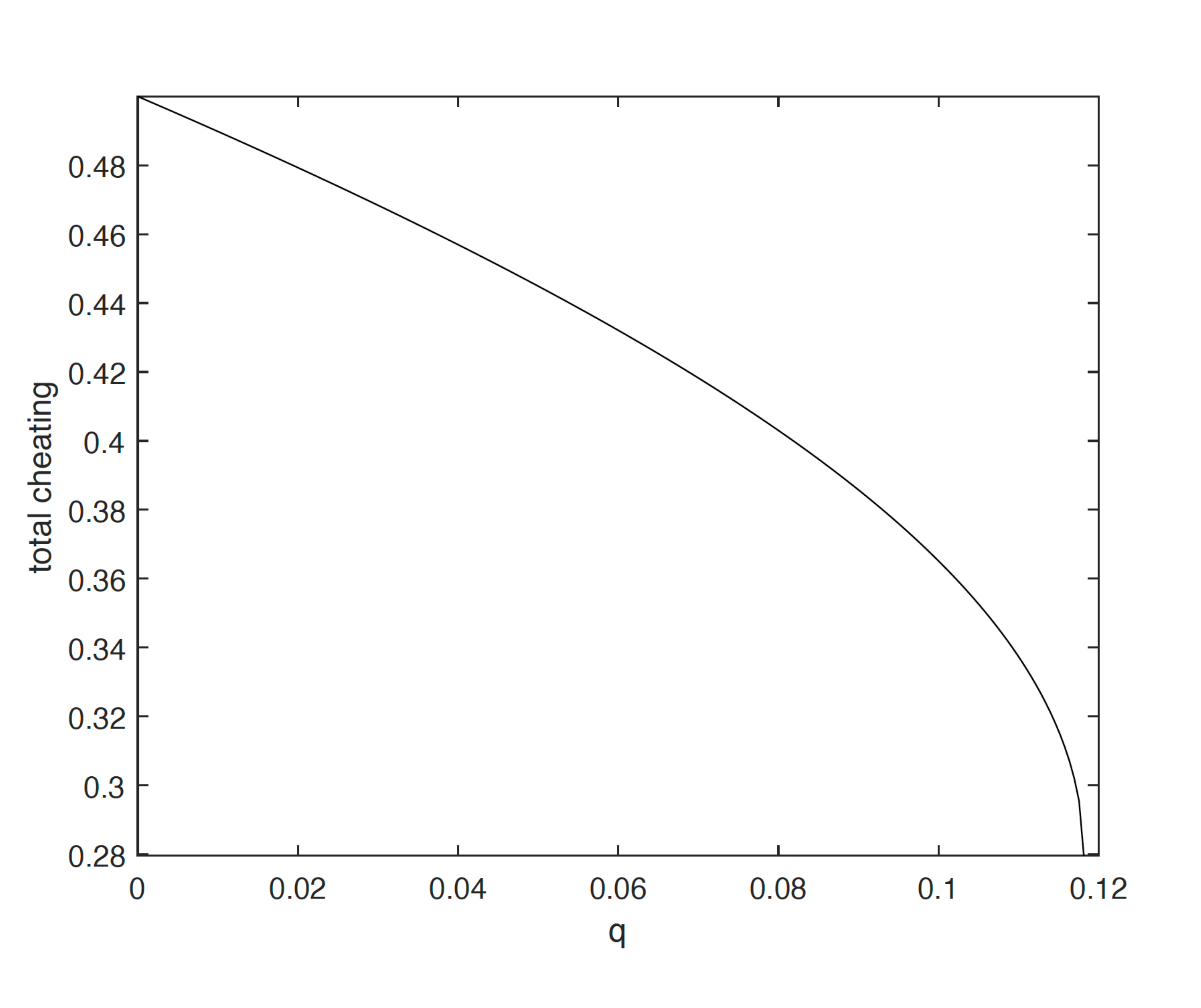}\end{center}
\vspace{-2.5em}
\caption{The total instance of cheating (vertical axis), by both scoundrels and responsives, in the bad 
equilibrium, as a function of the proportion of scoundrels $q$ (horizontal axis); for the case $\theta=1.5$.  As the proportion of scoundrels 
increases, total cheating diminishes, until the proportion of scoundrels nears $0.12$, at which point the bad and unstable 
equilibria vanish and only the good equilibrium remains.  At this point, the incidence of cheating drops from about $0.28$ to about $0.12$.
\vspace{-0.5em}
\label{sonata}}
\end{figure}

The total incidence of cheating in the bad equilibrium is $q+(1-q)s_b$.  Using \eqref{s_b} to substitute for $s_b$ and taking a derivative, a calculation shows that a society trapped at the bad equilibrium would  welcome more scoundrels.  The cheating of the additional scoundrels is overwhelmed by inducing responsives to cheat less and total cheating falls.  Eventually, the number of scoundrels increases to the point that only the good equilibrium remains.  Figure \ref{sonata} shows the total incidence of cheating as a function of the proportion of scoundrels, for a society with $\qqq=1.5$ that is trapped at the bad equilibrium (when the latter exists).  With no scoundrels, half of the agents cheat.  As the proportion of scoundrels increases, total cheating diminishes, until the proportion of scoundrels nears $12$ percent.  Here, the unstable equilibrium and the bad equilibrium coincide, a fraction of about $18$ percent of responsives cheat, and the total incidence of cheating, including the scoundrels, is about $28$ percent.   A further increase in the number of scoundrels then gives a discontinuous drop in the incidence of cheating, as society switches to the sole remaining equilibrium, the good one.

\section{Robustness to Invasion}\label{robot}

We now turn to the second of our resilience questions.  Beginning with a society that has settled on one of the stable equilibria, suppose the beliefs of a small fraction of the society's members are radically perturbed, in the form of an infusion of agents accustomed to the other stable equilibrium.  Will the original equilibrium survive, or will the infusion disrupt the standing equilibrium and prompt the society to converge to the other stable equilibrium? 

\subsection{Assimilation or Disruption?}\label{invite}

We continue to suppose that the social cost of cheating is sufficiently high ($\theta>1$) and there are sufficiently few scoundrels (i.e., $q<\hat q(\qqq)$) so that  we have three equilibria.  What happens when some outsiders, characterized by the behavior and perceptions of a society in the bad equilibrium, merge into a society characterized by the good equilibrium?  One can interpret this as a case in which a high-trust country (or organization, profession, culture, social group, and so on) is opened to entry 
(or membership, or participation, and so on) from agents accustomed to the bad equilibrium.   Will the newcomers be assimilated, and will their behavior converge
to that of the good equilibrium?  Or will the newcomers upset the social norm and cause everyone's behavior to settle on the bad equilibrium?

To address these questions, we suppose that a  population initially in the good equilibrium is shocked by the injection of a fraction $\lll$ $\leq$ $1/2$
of outsiders whose perception and behavior is taken from the bad equilibrium.   Refer to the members of the original population, who are now in proportion
$1-\lll$, as insiders and give them  subscript 1, and the invaders as outsiders, with subscript 0.  The basic equations for our system are then:%
\footnote{Implicit in this formulation is an assumption that insiders and outsiders  mix randomly.  We could alternatively imagine that outsiders are more likely to meet outsiders.}

\begin{eqnarray} \label{dyn_s}
\begin{array}{lcl}
	s&=& (1-\lll)^2\zzz_{11}+(1-\lll)\lll\zzz_{10}+\lll(1-\lll)\zzz_{01}+\lll^2\zzz_{00}\\
&&\\
	\zeta_{11} &=& \min\{\max\left\{0,x_1-f(s_{1})\right\},1\}\\
	\zeta_{10} &=& \min\{\max\left\{0,x_1-f(s_{0})\right\},1\}\\
	\zeta_{01} &=&\min\{ \max\left\{0,x_0-f(s_{1})\right\},1\}\\
	\zeta_{00} &=&\min\{ \max\left\{0,x_0-f(s_{0})\right\},1\}\\
	&&\\
	x_1 &=& \max\left\{f(s_1), \ds\frac12+\ds\frac12f(s_1)\right\}\\
	&& \\		
	x_0 &=& \max\left\{f(s_0), \ds\frac12+\ds\frac12f(s_0)\right\}~.
\end{array}
\end{eqnarray}
The variables $s_1$ and $s_0$ identify the proportion of cheating on the part of responsives perceived by insiders ($s_1$) and outsiders ($s_0$), and hence are the counterparts of $s_P$ from Section \ref{vitamin}.  As in Section \ref{vitamin}, we assume that these perceptions are commonly held by insiders and outsiders, reflecting their experience with their respective equilibria.   Insider and outsider proposers make offers that are optimal given their perceptions, and hence $x_1$ is the offer made by insiders and $x_0$ the offer made by outsiders.  Receivers make their decisions of whether to cheat based on the offer they face and their perception of the prevalence of cheating.

The proportion of responsives cheating in an interaction depends on both the identity of the proposer  and the identity of the responsive, and so we have four cheating probabilities to keep track of.  For example,  $\zzz_{10}$ is the proportion of cheating when an inside proposer interacts with an outside responsive.  The value of any $\zzz_{ij}$ can never go above 1 in equilibrium, so that the outer minimum in the specification of the four realizations of $\zeta_{ij}$ is redundant in equilibrium, but  $\zeta_{ij}$ can hit the upper bound of 1 in an out-of-equilibrium combination of a proposer who expects little cheating and hence makes a large offer with a responsive who expects a great deal of cheating and hence a low (social) cost of cheating.

The variable $s$ identifies the realized incidence of cheating among responsives.  
Each of the four terms corresponds to one of the four possible matches, involving either an inside or outside proposer and an inside or outside receiver, and gives the probability of such a match multiplied by the proportion of cheating in the match.

Again as in Section \ref{vitamin}, we assume that media reports, observations and informal communications prompt  the perceptions $s_1$ of insiders and $s_0$ of outsiders to both move toward the realization $s$, according to the dynamic system:%
\footnote{We specify the system directly in terms of the perceptions and realized cheating of responsives. An alternative but equivalent specification of the dynamic system would envisage the perception of \emph{total} cheating adjusting towards the realized  \emph{total} cheating: $\frac{d[(1-q)  s_k(t)+q]}{dt}= \ddd\{((1-q)s(t)+q) -((1-q)s_k(t)+q)\}$, with $k=0,1$.  Clearly, for any given $q$, this is equivalent to (\ref{dynamite}).}
\begin{eqnarray}\label{dynamite}
\begin{array}{lcr}
\dot s_1(t) & = & \ddd\{s(t) -s_1(t)\}\\\\
\dot s_0(t) & =& \ddd\{s(t) -s_0(t)\},
\end{array}
\end{eqnarray}
where $\delta>0$ allows us to tune the speed of adjustment.

\subsection{Convergence}

We first establish that perceptions converge.  The intuition is the following.
First, the dynamical system (\ref{dyn_s})---(\ref{dynamite}) implies that
\begin{eqnarray}\label{diffs1s0}
	s_0(t) = s_1(t) + e^{-\delta t}(s_0(0)-s_1(0)).
\end{eqnarray}
Therefore, the {\em difference} between $s_0(t)$ and $s_1(t)$ goes to zero as $t$ grows, i.e., the perceptions $s_1$ and $s_0$ of insiders and outsiders approach each other.
This is expected---both groups are adjusting their perceptions toward a common (though moving) level of realized cheating.  Second, once
these perceptions are sufficiently close, we essentially have the dynamic system described in Section \ref{vitamin} and pictured in Figure \ref{orchestra},
which converges to one of the two stable equilibria.  In Section \ref{sheepdog} we prove:

\begin{lemma}\label{conv2}
	The dynamical system (\ref{dyn_s})-(\ref{dynamite}) converges, with $\lim_{t\rightarrow\infty}s_1(t)=\lim_{t\rightarrow\infty}s_0(t)$
	and with both equal to either $s_g$, $s_u$, or $s_b$.	
\end{lemma}

\subsection{The Fragility of the Good Equilibrium}

When scoundrels are scarce, the good equilibrium is especially vulnerable to invasion.  If there are sufficiently few scoundrels, an arbitrarily small fraction $\lll$ of invaders from the bad equilibrium is capable of disrupting the good equilibrium: eventually all the agents converge to the beliefs and behavior of the bad equilibrium.  Section \ref{gadfly} in the Appendix proves:

\begin{proposition}\label{dvorak}
	Consider the dynamic system (\ref{dyn_s})-(\ref{dynamite}) with the initial conditions $s_1(0)=s_g$ $=$ $0$ and $s_0(0)=s_b$
	(i.e. a system in which the insiders initially believe themselves to be in the good equilibrium and outsiders in the bad equilibrium). For any $\lll >0$ there exists a
	$q^*>0$ such that, for any $q \leq q^*$ it will be the case that $\lim_{t\rightarrow\infty}s_1(t)=\lim_{t\rightarrow\infty}s_0(t)=s_b$, i.e. the
	system converges to the bad equilibrium.
\end{proposition}

When scoundrels are scarce, even a small influx of agents whose behavior initially matches the one prevailing in the bad equilibrium thus suffices to catapult the system into the basin of attraction of the bad equilibrium.  The basic intuition behind this result begins with the observation that as the proportion $q$ of scoundrels decreases, the basin of attraction of the good equilibrium  becomes smaller, as seen in Proposition \ref{sea}.  This alone does not explain the result, as the basins of attraction examined in Proposition \ref{sea} pertain to small shocks to the perception $s_P$ shared by all agents, whereas we are dealing here with a large shock to the perceptions of a small group of agents  The proof of Proposition \ref{dvorak} shows that nonetheless, when the proportion of scoundrels is small, such a small invasion has a large and quick enough effect on the perceptions of all agents as to pull the population away from the good equilibrium.  
 
 Proposition \ref{dvorak}  directs our attention to the fate of the good equilibrium in the face of small invasions.  Section \ref{paper} proves an expected monotonicity result for such invasions:

\begin{proposition}\label{halfpint}
[\ref{halfpint}.1]  If the good equilibrium survives an invasion of size $\lll\le 1/2$, it survives any invasion of size $\lll'<\lll$.  Similarly, if the good equilibrium is disrupted by an invasion of size $\lll<1/2$, it is disrupted by any invasion of size $\lll'\in[\lll,1/2]$.
	
\hspace{6em}[\ref{halfpint}.2] There is at most one value $\lll \in[0,1/2]$ such that an invasion of size $\lll$ gives convergence to the unstable equilibrium.
\end{proposition}	

\noindent  There are thus two possibilities. It may be that any invasion of size $\lll\le 1/2$ is unable to disrupt the good equilibrium.  This will be the case for relatively large values of $q$, i.e., when there are many scoundrels.  Alternatively, when $q$ is sufficiently small, the interval $[0,1/2]$ is partitioned by a value $\lll^*$, with smaller invasions being assimilated to the good equilibrium, invasions of size $\lll^*$ leading to the unstable equilibrium, and larger invasions disrupting the good equilibrium and leading to the bad equilibrium. 

\begin{figure}[H]
\vspace{-6em}
\hspace{1.5em}\includegraphics[scale=.357]{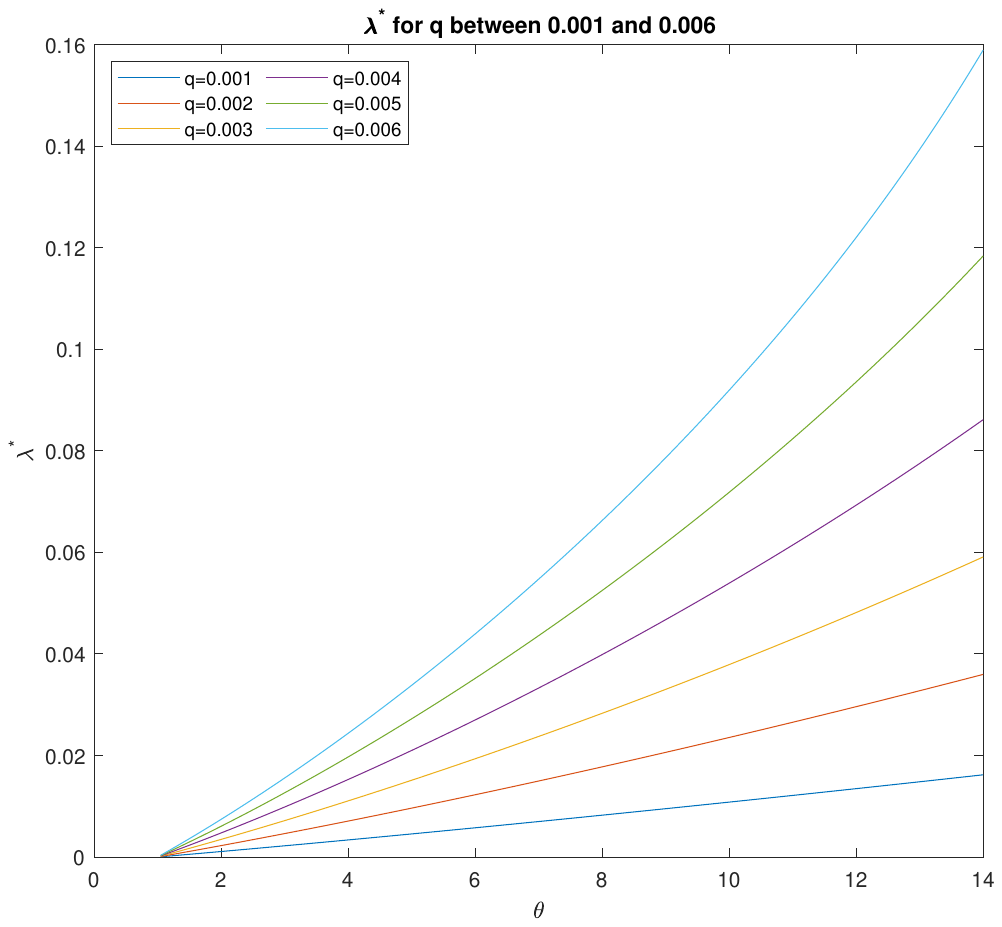}	\includegraphics[scale=.357]{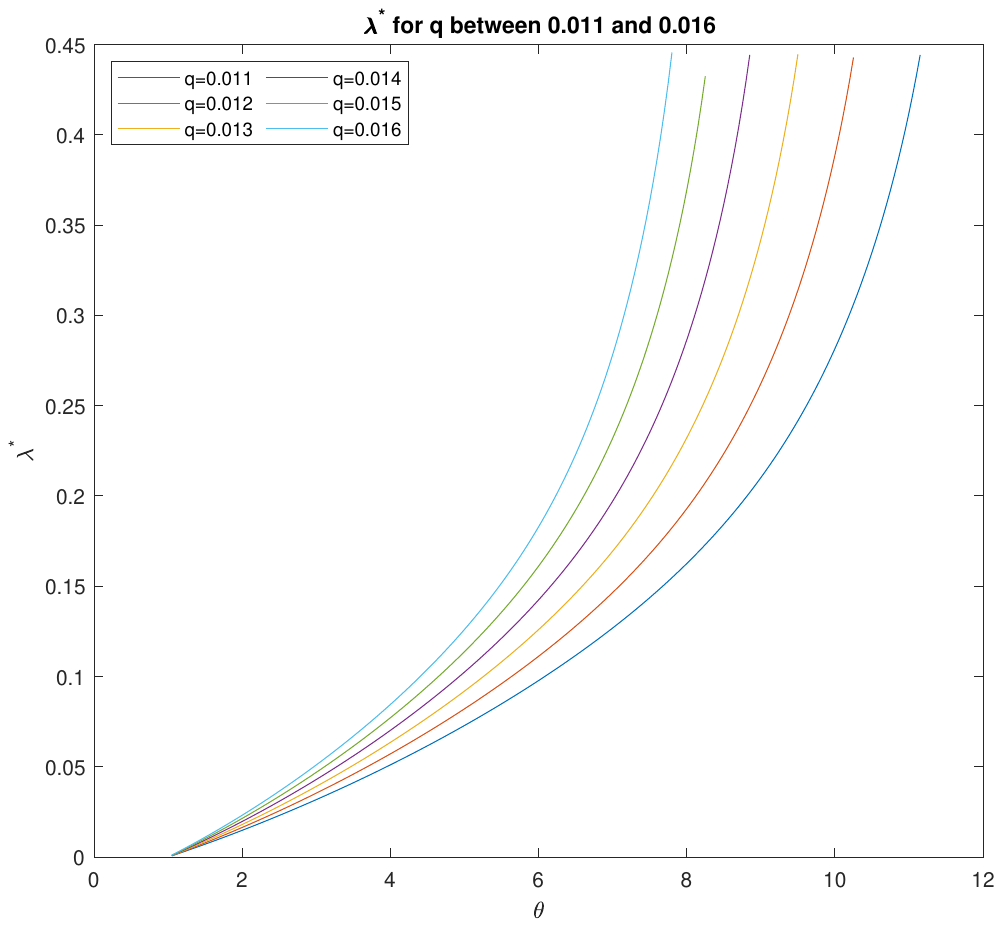}
\vspace{-6em}
\caption{Each panel shows, for the proportions $q$ of scoundrels indicated, the minimum invasion size $\lll^*$ (vertical axis) needed to disrupt the good equilibrium, as a function of $\theta$ (horizontal axis).  The good equilibrium can withstand larger incursions when $\theta$ is larger and when there are more scoundrels.  (Note the change in scale on the vertical axis in moving from panel (a) to (b).) Note also that when $q$ is larger, as in panel (b), the range of $\theta$ that are consistent with the presence of multiple equilibria shrinks, with smaller and smaller value of $\theta$ required to ensure that only the good equilibrium exists.
\label{merrimac}}
\end{figure}

Figure \ref{merrimac} shows the values of $\lll^*$ for selected values of $q$ 
and $\qqq$.\footnote{The value $\lll^*$ is computed as the value of $\lll$ such that, when its 15th decimal digit is reduced by 1, the limit to which the system converges shifts from the bad equilibrium to the good equilibrium.  This and the following figure are based on MatLab simulations of discrete approximations of our continuous dynamic system.}

%
%

%
%

\subsection{The Robustness of the Bad Equilibrium}

We  now turn this reasoning around. Consider the system (\ref{dyn_s})-(\ref{dynamite}) describing the dynamics of the perceptions of a mixture of agents, of which a fraction $(1-\lll)$ start with perception $s_g$ $=$ $0$ and a fraction $\lll$ start with perception $s_b$. If we now define $\hat{\lll}=1-\lll$, one readily sees that this is the same system as the one in which a proportion $1-\hat{\lll}$ of insiders whose  initial perception is $s_b$ and a
proportion $\hat{\lll}$ of  outsiders whose  initial perception is $s_g = 0$.  

Proposition \ref{dvorak} then gives:

\begin{corollary}\label{corollary: swap insiders and outsiders}
	Consider the dynamic system (\ref{dyn_s})-(\ref{dynamite}), assuming that for a fraction $1-\lll$ of insiders 
	$s_1(0)=s_b$ and for a fraction $\lll$ of outsiders $s_0(0)=s_g=0$ (i.e. a system in which the insiders initially 
	believe themselves to be in the bad equilibrium and the outsiders in the good equilibrium). For any $\lll <1$ there exists a $q^*>0$ 
	such that, for any\footnote{\rm Recall that we are interested in the case in which there are three equilibria. So 
	we must also have $q^* < \hat{q}(\theta)$ where $\hat{q}(\theta)$ is as in (\ref{mozart}).} 
	$q \leq q^*$ it will be the case that $\lim_{t\rightarrow\infty}s_1(t)=\lim_{t\rightarrow\infty}s_0(t)=s_b$, i.e. 
	the system converges back to the high cheating equilibrium.
\end{corollary}

The proof is almost immediate and can be found in Section \ref{section app: proof of corollary}.
The intuition mirrors that of Proposition \ref{dvorak}.  As  scoundrels become scarce, the basin of attraction of the bad equilibrium becomes large.  
It accordingly takes a large invasion of agents accustomed to the good equilibrium to disrupt the bad equilibrium.  In the extreme, as $q$ approaches zero, 
the basin of attraction of the bad equilibrium consumes the entire unit interval, allowing the bad equilibrium to withstand arbitrarily large invasions.  
Putting these results together, when scoundrels are scarce, the good equilibrium is upset by perturbations to bad behavior on the part of a tiny fraction of 
agents, while a large fraction of the population can shift to good behavior without disrupting the bad equilibrium.

\section{Discussion}\label{discuss}

Trust can be fragile.  When scoundrels are scarce, a high-trust equilibrium can be easily disrupted a small perturbation of the common perception of cheating (Proposition \ref{sea}) or by the injection of even a few bad apples (Proposition \ref{dvorak}), while a low-trust equilibrium can stubbornly resist the appearance of trusting agents.  In another version of the common saying, trust takes years to build, seconds to break, and forever to repair.

The basic forces behind these results are two-fold.  The possibility of multiple equilibria arises because the social cost of cheating is downward sloping---cheating is less costly when it is more prevalent---and nonlinear.%
\footnote{As Figure \ref{park} and Footnote \ref{coffee} indicate, either a convex or concave function $f$ can give rise to multiple equilibria.}
The relative stability properties of the equilibria, and in particular the relative fragility of the good equilibrium, arise because the cost of cheating is convex in the number of responsive cheaters, and decreases and becomes increasingly convex as the number of scoundrels decreases.  Together, these properties ensure that as scoundrels become scarce, the unstable equilibrium is pushed close to the good equilibrium while the bad equilibrium is pushed further away, shrinking the basin of attraction of the good equilibrium and increasing the basin of attraction of the bad equilibrium, as in Proposition \ref{sea}.  Intuitively, when scoundrels are scarce and cheating is low, it takes only a small increase in perceived cheating to sharply reduce the social cost, validating the increase and potentially catapulting the good equilibrium out of its (relatively small) basin of attraction.  

The convexity of the social cost of cheating also lies behind the relative susceptibility to invasion of the good equilibrium, a seen in Proposition \ref{dvorak}.  When scoundrels are scarce, agents in the high-trust equilibrium face a very steep portion of the cost function.  When an infusion of agents acclimated to the low-trust equilibrium raises the perceived level of cheating, the social falls sharply, inducing the formally high-trust agents to
cheat more, eventually pushing the society to the low-trust equilibrium. Conversely, agents
in the low-trust equilibrium face a much flatter portion of the cost curve. Hence, upon
observing less cheating than expected, their perceived social cost increases very little and
their cheating changes very little, allowing the low-trust equilibrium to survive.

The key properties of the social cost of cheating arise directly from the assumption that the social cost is directly proportional to the probability that a cheater is a scoundrel.  When very few responsives cheat, a cheater is almost certain to be a scoundrel, and hence cheating is punished heavily.  One readily notices and punishes as a likely scoundrel the only person who litters in a setting that everyone else preserves as pristine, or the only person who attempts
to jump a queue that everyone else scrupulously maintains, or the only person who breaks a traffic law that everyone else respects.  However, it initially takes only a modicum of cheating by responsives before a cheater is much less likely to be a scoundrel, 
and so the cost of cheating initially drops very rapidly as the incidence of cheating increases.   When cheating is rampant, a transgressor is less likely to be a scoundrel and so punished less heavily.  Moreover, an increase in the incidence of cheating has little effect on the likelihood a cheater is a scoundrel, and so the cost of cheating falls less and less rapidly as the incidence of cheating increases.  
The fewer the scoundrels, the more pronounced the effect of having even a few responsives among the ranks of cheaters, and so the more pronounced this convexity.

\begin{figure}[H]
	\begin{center}
		\includegraphics[scale=.75]{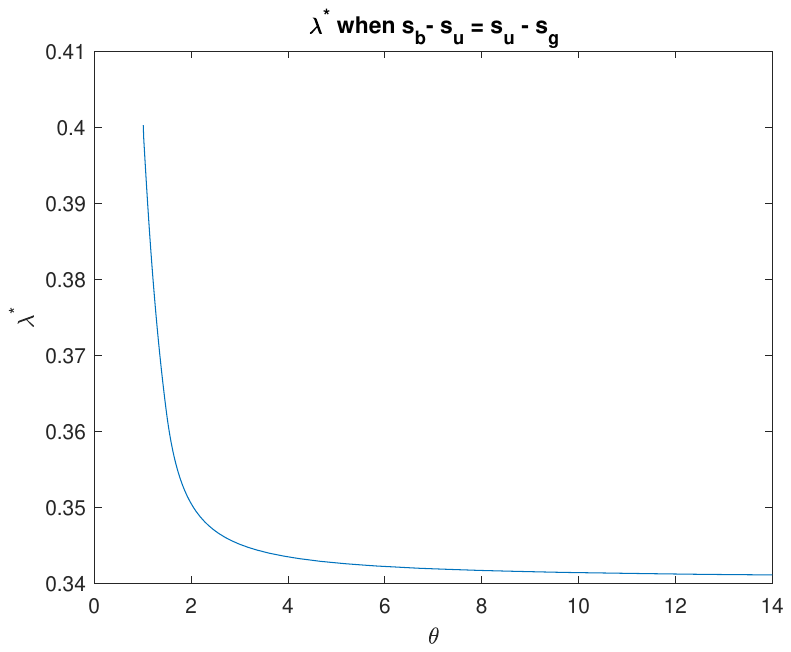}	
		\caption{For each value of the cheating-cost parameter $\theta$, the proportion of scoundrels $q$ is set 
			so that the unstable equilibrium is halfway between the good and the bad equilibria.  We then numerically calculate $\lll^*$, 
			the size of infusion of agents from the bad equilibrium just sufficient to disrupt the good equilibrium. As $\theta$ 
			increases above 1, the proportion of scoundrels required to keep the unstable equilibrium halfway between the other two equilibria decreases, making the cost-of-cheating function more convex, and hence reducing the infusion 
			of agents accustomed to the bad equilibrium  that suffices to disrupt the good equilibrium.	\label{hippo}}
	\end{center}
	\vspace{-2em}
\end{figure}

Section \ref{robot} showed that when there are few scoundrels, an arbitrarily small invasion of agents accustomed to the bad equilibrium can disrupt the good equilibrium.   One might think that this is nothing more than a manifestation of Section \ref{vitamin}'s result that the basin of attraction of the good equilibrium is small.   To see that this is not the case, Figure \ref{hippo} reports results of the following exercise.
For various values of the cheating-cost parameter $\theta$, we set the proportion of scoundrels $q$ so that the unstable equilibrium $s_u$ is halfway between the good and bad equilibria. Since $s_u$ is the common boundary of the basins of attraction of the two stable equilibria, by keeping it always exactly in their middle we make sure that the distance that the system needs to travel before being drawn towards the bad equilibrium, starting from the good one, is equal to the distance that needs to be traveled before being drawn towards the good equilibrium, starting from the bad one.\footnote{Injecting a small fraction of agents accustomed to one equilibrium into a society sitting in the other equilibrium implies that the system will typically move between the two equilibria, making the portion of the basin of attraction of the bad equilibrium between $s_b$ and 1 irrelevant.} 
We then numerically calculate $\lll^*$, the size of infusion of agents from the bad equilibrium 
just sufficient to disrupt the good equilibrium, for each of these cases.  

Over the relevant range the value of $\lll^*$ is always clearly below $0.5$ (it is in fact below $0.4$).  This, using Propositions \ref{dvorak} and Corollary \ref{corollary: swap insiders and outsiders}, implies that  more outsiders are needed to disrupt the bad equilibrium than are needed to disrupt the good equilibrium, even when the dislocation of perceptions needed to push agents accustomed to the good equilibrium into the basin of attraction of the bad equilibrium is equal to the dislocation needed to push agents accustomed to the bad equilibrium into the basin of attraction of the good one. As $\theta$ increases, the proportion of scoundrels required to maintain $s_u$ at the mid-point between $s_g$ and $s_b$ decreases, making the cost-of-cheating function more convex, and hence reducing the infusion of agents 
accustomed to the bad equilibrium  that suffices to disrupt the good equilibrium.

\begin{figure}[H]
	\hspace{2em}
\hspace{-3.5em} \includegraphics[scale=.25]{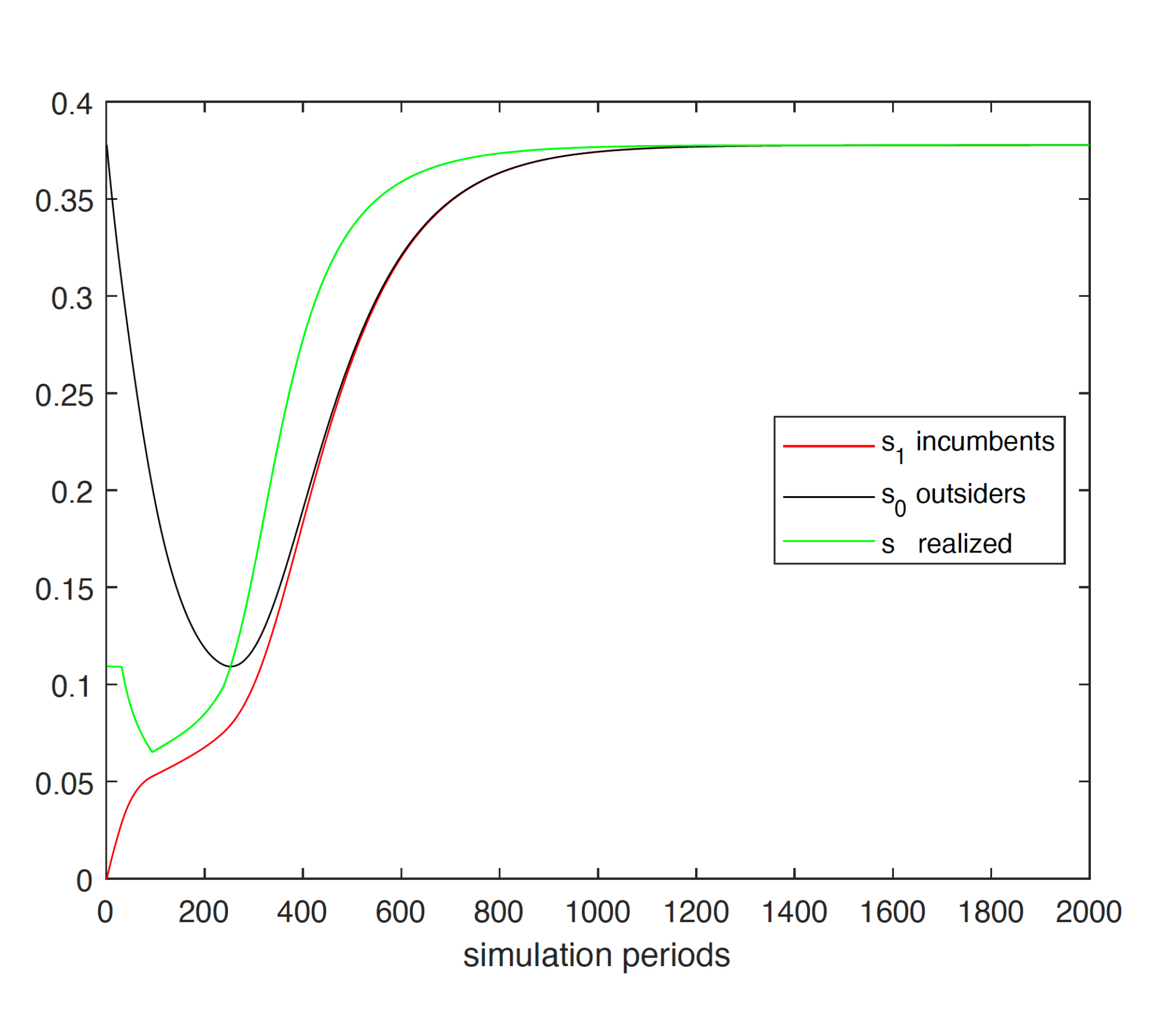} \vspace{5em}\includegraphics[scale=.25]{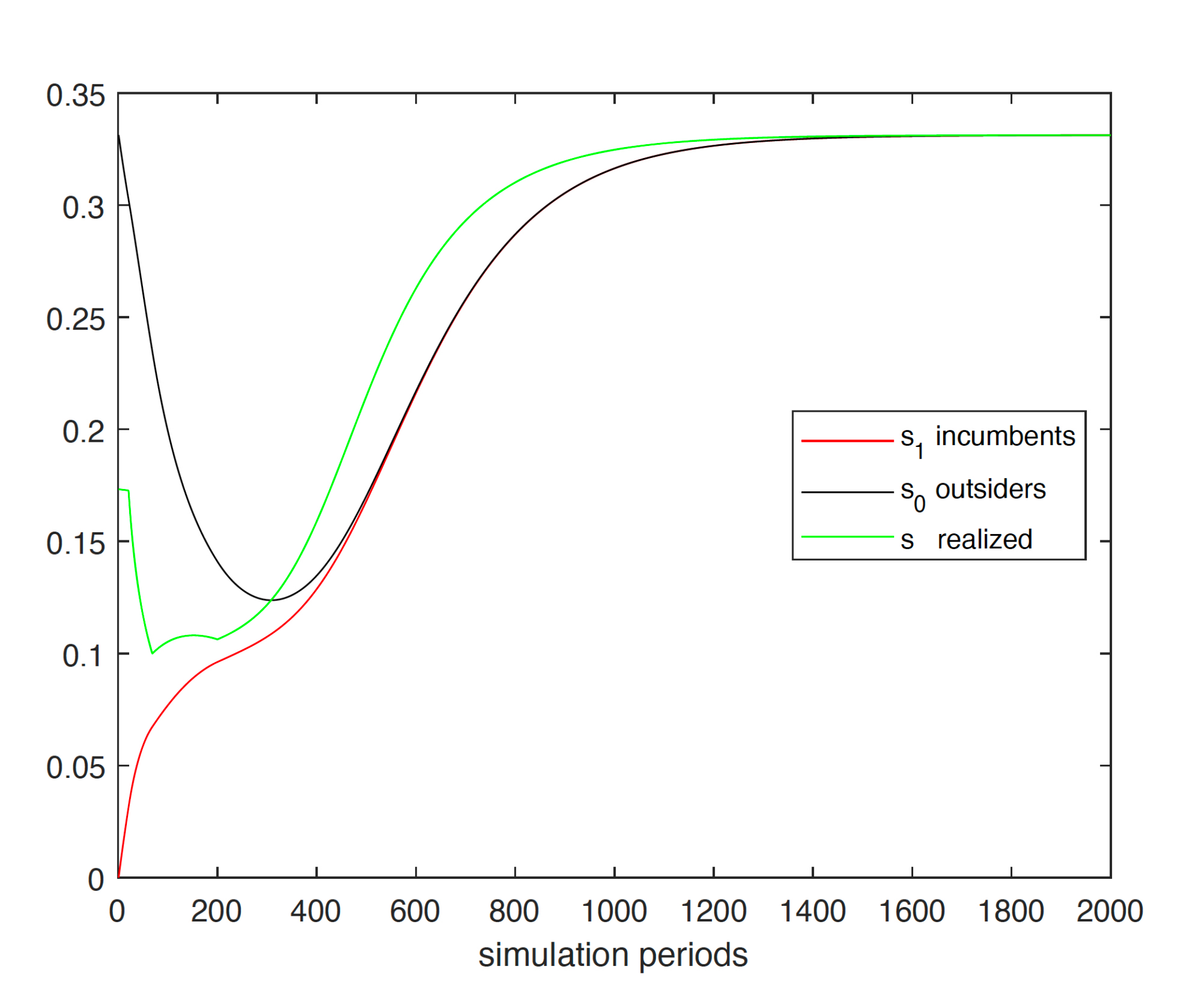}	
\vspace{-6em}	\caption{Depiction of the dynamics for two cases in which an invasion of agents 
accustomed to the
		bad equilibrium leads the population to converge to the bad equilibrium.  Each panel shows the 
		paths of the insider perception of cheating $s_1$ (red), outsider perception of cheating $s_0$ (black) and 
		the realized perceptions of cheating $s$ (green). The parameters underlying the left panel are $\theta=2, \lll=0.118, q=0.05$, those in the right panel are 
 $\theta=2, \lll=0.2, q=0.0634$. The  
		kinks in the paths arise as various of the min and max operators in \eqref{dyn_s} come into play.\label{camel}}
\end{figure}

Figure \ref{camel} portrays the path of the dynamic systems, for two cases in which an invasion disrupts 
the good equilibrium and induces convergence to the bad equilibrium.  In each case,  $s_1$ initially equals $s_g$ $=$ $0$ 
(insider perception and behavior are initially consistent with the good equilibrium) and $s_0$
initially equals $s_b$ (outsider perception and behavior are initially consistent with the bad equilibrium).

Two aspects of these dynamics stand out.  First, $s_1$ initially increases as insiders adjust to the more-than-expected 
cheating carried out by outsiders.  However, the outsiders' perceived level of cheating $s_0$ falls, as they meet less cheating 
than expected when matched with insiders.  The outsiders' perceived level of cheating falls much more dramatically, reflecting 
their smaller share of the population, and hence the realized level of cheating, $s$, on balance falls (only imperceptibly at the beginning).  There thus initially appears to
 be overwhelming evidence that the population is adjusting toward the good equilibrium.  However, in both cases the 
 direction of $s$ eventually reverses (after some seeming indecision in the right panel) and the population converges
to the bad equilibrium.  Second, the adjustment of the aggregate level of cheating not only need not be monotonic, 
but can be complicated, in the right panel reversing direction three times.

The idea that trust can be fragile is familiar.  The more surprising finding to emerge from this exploration is that, perhaps paradoxically, trust can be more robust when there are more agents in the economy who can never be trusted. Intuitively, this is because social disapproval is heaped on cheaters who do so without a good reason. The more reckless cheaters there are\textemdash the more scoundrels, as we call them\textemdash the more likely a person observed cheating is one of them, and therefore the more his cheating is socially sanctioned. 
Scoundrels can thus be valuable for two reasons.  Increasing the number of scoundrels may convert an economy with multiple equilibria into an economy with a unique (good) equilibrium.  As we have seen in Section \ref{vitamin}, if the former economy is coordinated on the less trusting of the multiple equilibria, the increase in scoundrels leads to an increase in trust. In addition, an increase in the number of scoundrels can render an economy coordinated on the 
most trusting of multiple equilibria better able to withstand perturbations to that equilibrium.  The most fortunate economy is one that has few scoundrels, and hence multiple equilibria,  but that has coordinated on the high-trust equilibrium.   
But the  higher is the level of trust in the good equilibrium (i.e., the fewer scoundrels), the more precarious is the equilibrium itself.

Our analysis points to steps that might mitigate this fragility.  If we broadened the purview of our analysis to accommodate either multiple or continuous arrivals of outsiders, then we expect that an economy whose good equilibrium would be disrupted by a moderate influx of outsiders accustomed to the bad equilibrium could accommodate an even larger number of such additions if they occur sufficiently slowly.  This moderated flow would allow previous arrivals to have time to adjust and thus keep the system within the basin of attraction of the good equilibrium, even as the flow of new arrivals continues.  Taking steps to hasten the adjustment of perceptions would allow the good equilibrium to withstand a larger influx of outsiders, but taking steps to reduce the number of scoundrels would have the reverse effect.   We can expect an institution devoid of scoundrels (perhaps Minnesota?)  to have more difficulty accommodating arrivals accustomed to the bad equilibrium than a somewhat grittier one (perhaps New York?).

We have worked throughout with the simple specification of the social cost of cheating given by \eqref{brahms} and simple, symmetric adjustment dynamic given by \eqref{dynamite}.  We believe that, if anything, the more realistic components we might build into these specifications would reinforce our basic finding that trust is likely to be fragile.  For example, we expect violations of trust in a high-trust environment to be more visible and more salient than episodes of trust in a low-trust environment.  If so, the tendency of shocks to disrupt a high-trust equilibrium will be exacerbated.  


\newpage{}

\fontsize{12}{13.89}\selectfont

\newpage{}

\begin{appendix}

\begin{center}
{\bf For Online Publication}
\end{center}

\section{Proof of Proposition \ref{cantata}}\label{mahler}

From \eqref{s star defined}, if $\qqq<1$, then $s^*<0$. Hence the equilibrium conditions \eqref{eric}--\eqref{ginger} reduce to
\begin{eqnarray*}
	s &= &\max\{0,x-f(s)\}\\
	&=& \max\left\{0,\frac12-\frac12f(s)\right\}\\
	&=&\max\left\{0,\frac12-\frac12\frac{\qqq q}{q+(1-q)s}\right\}\\
	&=&\frac12-\frac12\frac{\qqq q}{q+(1-q)s}.
\end{eqnarray*}
Given $\qqq<1$, this equation has only one positive (real) solution.\hfill\endproof

\section{Proof of Proposition \ref{sea}}\label{app: proof of prop Sea} 
Straightforward manipulations of \eqref{s_b} and \eqref{s_u}, in the range consistent with $q<\hat q(\qqq)$, imply that $s_u$ is an increasing function of $q$ and $\qqq$, while $s_b$ is a decreasing function of $q$ and $\qqq$, with
\begin{eqnarray}\nonumber
	\lim_{q\rightarrow 0}s_u(q) \; = \;  0 \qquad {\rm and} \qquad	
	\lim_{q\rightarrow 0}s_b(q) \; = \; \frac12.
\end{eqnarray}	
\hfill\endproof

\section{Proof of Lemma \ref{conv2}}\label{sheepdog}

We first note that
for $s_u \leq s \leq s_b$, we have that
\begin{eqnarray}\label{graphin}
	\frac12 -\frac12 f(s) \geq s
\end{eqnarray}
with a strict inequality except at the two boundaries, while for both $s<s_u$ and $s>s_b$, it is true that
\begin{eqnarray}\label{graphout}
	\frac12 -\frac12 f(s) < s.
\end{eqnarray}

We can write  $s(t) = h(s_0(t),s_1(t))$ and then write the dynamical system (\ref{dyn_s})-(\ref{dynamite}) as
\begin{eqnarray}\nonumber
	\dot s_1(t)= \ddd\{h(s_0(t),s_1(t)) -s_1(t)\}\\
	\dot s_0(t) = \ddd\{h(s_0(t),s_1(t)) -s_0(t)\},
\end{eqnarray}
where the function $h(s_0(t),s_1(t))$ is derived from \eqref{dyn_s} and  gives the realized proportion of cheating by responsives, 
$s(t)$, as a function of the current state of the perceptions by outsiders and incumbents, respectively, $(s_0(t),s_1(t))$.  In the following argument, 
we repeatedly use the facts that  the function $h$ is uniformly continuous on $[0,1]^2$, and that along the diagonal $s_1(t)=s_0(t)=s$, the function $h$ is given by
\begin{eqnarray}\nonumber
h(s,s) =
\left\{
\begin{array}{lll}
	0&~~~&s\le s^*\\
	\displaystyle\frac12-\frac12f(s)&&s\ge s^*,
\end{array}
\right.
\end{eqnarray}
and hence, as implied by (\ref{graphin}) and (\ref{graphout}), we have
\begin{eqnarray}\nonumber
	h(s,s)-s <0&~~~&s<s_u\\
	h(s,s)-s =0&~~~&s=s_u\\
	h(s,s)-s >0&~~~&s_u<s<s_b\\
	h(s,s)-s =0&~~~&s=s_b\\
	h(s,s)-s <0&~~~&s>s_b.
\end{eqnarray}	

Fix a sufficiently small $\eta>0$.  Then there exists $\eee(\eta)>0$ such that
\begin{eqnarray}
s\in[\eta,s_u-\eta] &\implies& h(s,s)-s<-\eee(\eta)\label{tinker}\\
s\in[s_u+\eta,s_b-\eta] &\implies& h(s,s)-s>~~\eee(\eta)\label{evers}\\
s\in[s_b+\eta,1] &\implies& h(s,s)-s<-\eee(\eta).\label{chance}
\end{eqnarray}
Let $|\!|\cdot|\!|$ denote the sup norm. %
There exists $\ggg(\eta)>0$ sufficiently small such that $|\!|(s_0,s_1)-(s_0,s_0)|\!|<\ggg(\eta)$ implies%
\footnote{The first inequality follows from the absolute continuity of $h$.  We can ensure the second by taking $\gamma(\eta)$ to be sufficiently small.}
\begin{eqnarray}\nonumber
|h(s_0,s_1)-h(s_0,s_0)|&<&\frac{\eee(\eta)}{4}\\
|s_0-s_1|&<&\frac{\eee(\eta)}{4},
\end{eqnarray}
which in turn imply, using the triangle inequality,
\begin{eqnarray}\label{abbott}
	|h(s_0,s_1)-s_0)-(h(s_0,s_0)-s_0)|&<&\frac{\eee(\eta)}{2}
\end{eqnarray}
\begin{eqnarray}\label{costello}
	|(h(s_0,s_1)-s_1)-(h(s_0,s_0)-s_0)|&<&\frac{\eee(\eta)}{2}.
\end{eqnarray}
Hence, whenever $|\!|(s_0,s_1)-(s_0,s_0)|\!|<\ggg(\eta)$, we can combine \eqref{abbott} and \eqref{costello} 
with \eqref{tinker}--\eqref{chance}, to establish the following implications:%
\begin{eqnarray}
	s_0,s_1\in[\eta,s_u-\eta] & \implies &
\left[h(s_0,s_1)-s_0<-\frac{\eee(\eta)}{2},~~~h(s_0,s_1)-s_1<-\frac{\eee(\eta)}{2}\right] \label{beck}
\end{eqnarray}
\begin{eqnarray}
\hspace{-3em}	s_0,s_1\in[s_u+\eta,s_b-\eta] & \implies &
\left[h(s_0,s_1)-s_0>\frac{\eee(\eta)}{2},~~~h(s_0,s_1)-s_1>~~\frac{\eee(\eta)}{2}\right]\label{bogart}
\end{eqnarray}
\begin{eqnarray}	
	s_0,s_1\in[s_b+\eta,1] & \implies &
\left[h(s_0,s_1)-s_0<-\frac{\eee(\eta)}{2},~~~h(s_0,s_1)-s_1<-\frac{\eee(\eta)}{2}\right].\label{appice}
\end{eqnarray}
From \eqref{diffs1s0}, we see that there exists $T(\eta)$ such that for all $t>T(\eta)$, we have
$|\!|(s_0,s_1)-(s_0,s_0)|\!|<\min\{\eta,\ggg(\eta)\}$.  The preceding three implications then imply two possibilities:

\begin{itemize}
	\item For all $t>T(\eta)$, $s_0$ and $s_1$ are both within $2\eta$ of $s_u$.
	
	\item There is a time $t'>T$ at which at least one of $s_0$ or $s_1$ differ from $s_u$ by more than $2\eta$.  
	Then both $s_0$ and $s_1$ differ from $s_u$ by more than $\eta$. %
	 Hence, \eqref{beck}--\eqref{appice} imply that there exists a time $t''\ge t'$ such that for all $t>t''$,
	either both $s_0$ and $s_1$ differ from $s_g$ by at most $2\eta$ (from \eqref{beck}) or both $s_0$ and $s_1$ differ from $s_b$ by at most $2\eta$  (from \eqref{bogart}--\eqref{appice}).
\end{itemize}

Since this holds for any $\eta>0$, we have convergence. \hfill\endproof

\section{Proof of Proposition \ref{dvorak}}\label{gadfly}

We begin with a preliminary result.

\begin{lemma}\label{lemma app: go to sb after su}
	Let $q<\hat{q}(\qqq)$, so that there are 3 distinct equilibria. If at some finite time $t$ it is the case 
	that $s_1(t)=s_u$, the dynamic system (\ref{dyn_s})-(\ref{dynamite}), with initial conditions $s_1(0)=0$ and $s_0(0)=s_b$, converges to $s_b$.
\end{lemma}	

\proof Using (\ref{diffs1s0}) we can write the dynamics entirely in terms of $s_1(t)$ and $t$, for a given $\lll$:
\begin{eqnarray}\label{dyn_gen}
\begin{array}{l}
	\dot s_1(t) =  \\
	\\
	\ddd \left\{
	(1-\lll)^2 \min\{1,\max\{0,\max\{f(s_1(t)), \ds\frac12+\ds\frac12f(s_1(t))\} 
	-f(s_1(t))\}\}\right.\\ 
	 \\
	+\lll(1-\lll) \min\{1,\max\{0,\max\{f(s_1(t)), \ds\frac12+\ds\frac12f(s_1(t))\} 
	-f(s_1(t)+e^{-\ddd t}s_b)\}\}\\ 
	 \\
	+\lll(1-\lll) \min\{1,\max\{0,\max\{f(s_1(t)+e^{-\ddd t}s_b), \ds\frac12+\ds\frac12f(s_1(t)+e^{-\ddd t}s_b)\}
	-f(s_1(t))\}\}\\
	 \\
	\left.+ \lll^2 \min\{1,\max\{0,\max\{f(s_1(t)+e^{-\ddd t}s_b), \ds\frac12+\ds\frac12f(s_1(t)+e^{-\ddd t}s_b)\}
	-f(s_1(t)+e^{-\ddd t}s_b)\}\}- s_1(t)\right\}.
\end{array}
\end{eqnarray}
Assume now that, at some finite $t$, $s_1(t)\geq s_u$. For any $\qqq>1$ this implies that
\[
s_0(t)>s_1(t)\geq s_u>s^*>0.
\]
As long as $s_1(t)<s_b$ (which is strictly larger than $s_u$, given that $q<\hat{q}(\qqq)$), we can simplify the dynamics, 
since all the inner max appearing in (\ref{dyn_gen}) are solved by the second of the two expressions.
%
More in detail, in the expression multiplied by $(1-\lll)^2$ we have:
\begin{eqnarray}\nonumber
	\max\left\{f(s_1(t)), \ds\frac12+\ds\frac12f(s_1(t))\right\}-f(s_1(t))\; = \;
	\ds\frac12-\ds\frac12f(s_1(t)).
\end{eqnarray}
In the first of the two expressions multiplied by $\lll(1-\lll)$ we have:
\begin{eqnarray}\nonumber
	\max\left\{f(s_1(t)), \ds\frac12+\ds\frac12f(s_1(t))\right\}-f(s_1(t)+e^{-\ddd t}s_b)=
	\ds\frac12+\ds\frac12f(s_1(t))-f(s_1(t)+e^{-\ddd t}s_b).
\end{eqnarray}
In the second of the two expressions multiplied by $\lll(1-\lll)$ we have:
\begin{eqnarray}\nonumber
	\max\left\{f(s_1(t)+e^{-\ddd t}s_b), \frac12+\ds\frac12f(s_1(t)+e^{-\ddd t}s_b)\right\}-f(s_1(t))=
		\ds\frac12+\ds\frac12f(s_1(t)+e^{-\ddd t}s_b)-f(s_1(t)).
\end{eqnarray}
Note that 
\begin{eqnarray}\nonumber
\ds\frac12+\ds\frac12f(s_1(t))-f(s_1(t)+e^{-\ddd t}s_b)>0
\end{eqnarray}
and since $s_1(t)>s^*$ this implies that 
\begin{eqnarray}\nonumber
1>f(s_1(t))>f(s_1(t)+e^{-\ddd t}s_b).
\end{eqnarray}
Therefore, the first of the two expressions multiplied by $\lll(1-\lll)$ reduces to  
\begin{eqnarray}\nonumber
\frac12+\frac12f(s_1(t))-f(s_1(t)+e^{-\ddd t}s_b).
\end{eqnarray}
Since 
\begin{eqnarray}\nonumber
\frac12+\frac12f(s_1(t)+e^{-\ddd t}s_b)-f(s_1(t))
\end{eqnarray}
cannot be signed, the second of the two expressions in (\ref{dyn_gen}) multiplied 
by $\lll(1-\lll)$ only reduces to 
\begin{eqnarray}\nonumber
\max\left\{0,\frac12+\frac12f(s_1(t)+e^{-\ddd t}s_b)-f(s_1(t))\right\}.
\end{eqnarray}
Finally, for the expression in (\ref{dyn_gen}) multiplied by $\lll^2$ we have:
\begin{eqnarray}\nonumber
	\max\{f(s_1(t)+e^{-\ddd t}s_b), \frac12+\frac12f(s_1(t)+e^{-\ddd t}s_b)\}-f(s_1(t)+e^{-\ddd t}s_b)=
		\frac12-\frac12f(s_1(t)+e^{-\delta t}s_b).
\end{eqnarray}
Putting together all these observations about the four components of the right side of (\ref{dyn_gen}) we get
\begin{eqnarray}\label{dyns3}
\begin{array}{l}
	\dot s_1(t) =  \\
	\\
	\ddd \left\{
	(1-\lll)^2\left(\ds\frac12-\ds\frac12f(s_1(t)\right)\right.\\ 
	 \\
	+\lambda(1-\lambda)\left(\ds\frac12+\ds\frac12f(s_1(t))-f(s_1(t)+e^{-\ddd t}s_b)+
    \max\left\{0,\ds\frac12+\ds\frac12f(s_1(t)+e^{-\ddd t}s_b)-f(s_1(t))\right\}\right)\\ 	 
	 \\
	\left.+ \lll^2\left(\ds\frac12-\ds\frac12f(s_1(t)+e^{-\delta t}s_b)\right)-s_1(t)\right\}\geq \\
	\\
	\hfill\ddd \left\{(1-\lll)^2\left(\ds\frac12-\ds\frac12f(s_1(t))\right)+\lll(1-\lll)\left(1-\ds\frac12f(s_1(t))-\ds\frac12f(s_1(t)+e^{-\ddd t}s_b)\right)+\right.\\
	\\
	\hfill\left.\lll^2\left(\ds\frac12-\ds\frac12f(s_1(t)+e^{-\ddd t}s_b)\right)-s_1(t)\right\}  \\
	\\
	\hfill=\ddd \left\{\ds\frac12-\ds\frac12f(s_1(t))+\ds\frac{\lll}{2}(f(s_1(t))-f(s_1(t)+e^{-\ddd t}s_b))-s_1(t)\right\},
\end{array}
\end{eqnarray}
where the middle inequality results from neglecting the $\max$ operator.

Given that $s_u\leq s_1(t)<s_b$, we know that 
\begin{eqnarray}\nonumber
s_1(t)\leq\ds\frac12-\ds\frac12f(s_1(t)) \;\; \Leftrightarrow \;\; \ds\frac12-\ds\frac12f(s_1(t))-s_1(t)\geq 0.
\end{eqnarray}
Moreover, since $f$ is decreasing, for any finite $t$ we have
\begin{eqnarray}\nonumber
f(s_1(t))-f(s_1(t)+e^{-\ddd t}s_b)>0
\end{eqnarray}

Hence $\dot s_1(t)>0$ for all $s_u\leq s_1(t)<s_b$.  Since we know that the system converges, it must then be that $s_1(t)$ converges 
to $s_b$. \hfill\endproof

\bigskip

The proof of Proposition \ref{dvorak} now proceeds in four steps.

\begin{step}{Bounding $s_0$ from below for an initial interval of time}\end{step}

First, fix $\lll \leq 1/2$, $\theta$
and a value of $0<q<\hat{q}(\qqq)$, to guarantee that there are three equilibria (to simplify the notation, we will denote this as $\hat{q}$).  Recall the dynamics
\begin{eqnarray}\nonumber
	\dot s_1&=& \ddd(s-s_1)\\
	\dot s_0 &=& \ddd(s-s_0).
\end{eqnarray}

Recall that $s_{00}$ is the  amount of cheating that occurs when an outsider proposer meets an outside receiver.  At time $0$, we have $s_{00} = s_b$, where we recall that the latter is the level of cheating characterizing the bad equilibrium.  Then in general we have, using (\ref{dyn_s}),
\[
s\ge \lll^2s_{00},
\]
and hence
\begin{eqnarray}\nonumber
	\dot s_1&\ge& \ddd(\lll^2s_{00}-s_1)\\
	\dot s_0 &\ge& \ddd(\lll^2s_{00}-s_0).
\end{eqnarray}
Now we note that, as long as $s_0>s^*$(which initially must be the case given that $s_0(0)=s_b>s^*$), we have
\[
s_{00} = \frac12-\frac12\frac{\qqq q}{q+(1-q)s_0},
\]
and so we can write
\begin{eqnarray}\nonumber
	\dot s_1&\ge& \ddd\left(\lll^2\left( \frac12-\frac12\frac{\qqq q}{q+(1-q)s_0}\right)-s_1\right)\label{dynLB1}\\
	\dot s_0 &\ge& \ddd\left(\lll^2\left( \frac12-\frac12\frac{\qqq q}{q+(1-q)s_0}\right)-s_0\right).\label{dynLB2}
\end{eqnarray}
The right hand side in (\ref{dynLB2}) is larger than the expression we obtain by setting to 0 the $s_0$ that appears in the denominator. Hence we have
%
\begin{eqnarray}\nonumber
\dot s_0 \ge \ddd\left(\lll^2\left( \frac12-\frac12\qqq \right)-s_0\right)
\end{eqnarray}
for all $q\in (0,\hat q)$.

Hence, for any $\eta>0$, there exists a time $t_{\eta}>0$ such that $s_0(t)\ge s_b-\eta$ for all $t\in [0,t_{\eta}]$.

\begin{step}{Bounding $s_1$ from below at a given point in time}\end{step}
\bigskip

Consider now (\ref{dynLB1}). The expression within the inner brackets is increasing in $s_0$ and decreasing in $q$.
Therefore, over the interval $[0,t_{\eta}]$, replacing $s_0$ by its lower bound of $s_b-\eta$, and again $s_b$ by its lower bound\footnote{See
(\ref{s_b}) and (\ref{mozart}).}
of $(1-3\hat q)/(4(1-\hat q))$,
we reduce that expression. We also reduce it replacing $q$ by its upper bound of $\hat q$. Combining these changes we obtain a 
lower bound on the right side of (\ref{dynLB1}) that implies
\begin{eqnarray}\nonumber
\dot s_1(t)\ge \ddd\left(\lll^2\left( \frac12-\frac12\frac{\qqq \hat q}
{\hat q(\frac{1}{4}+\eta)+(\frac{1}{4}-\eta)}\right)-s_1(t)\right).
\end{eqnarray}
It is a bit tedious but straightforward to verify that, for any $\qqq>1$ it must be that
\begin{eqnarray}\nonumber
\ds\frac12\left(1 - \ds\frac{\qqq \hat q}
{\ds\frac{1}{4}(\hat q+1)}\right)>0.
\end{eqnarray}
We can then choose $\eta$ sufficiently small so that
\begin{eqnarray}\nonumber
\frac12 \left(1 - \frac{\qqq \hat q}
{\hat q(\ds\frac{1}{4}+\eta)+(\ds\frac{1}{4}-\eta)}\right)>0.
\end{eqnarray}
Then we have that
\begin{eqnarray}\nonumber
\dot s_1(t) \ge \delta(A-s_1(t))
\end{eqnarray}
for some $A>0$ and for any fixed $q\in (0,\hat q)$ and all $t\in [0,t_{\eta}]$.

Hence, there exists a time $\tau\in[0,t_{\eta}]$ and value $\xi>0$ such that, for any fixed $q\in (0,\hat q)$,  we have,
\begin{eqnarray}\nonumber
s_1(\tau)\ge\xi>0.
\end{eqnarray}

\begin{step}{Pushing $s_u$ below $s_1$.}\end{step}

Now let $q$ approach 0.  As we do so, $s_u(q)\rightarrow 0$.  Hence, for all sufficiently small $q$, at time $\tau$ we have $s_1(\tau)>s_u$.

\begin{step}{Showing convergence to $s_b$.}\end{step}
\bigskip

We can now invoke Lemma \ref{lemma app: go to sb after su} and conclude that $s_1(t)$ converges to $s_b$.\hfill\endproof

\section{Proof of Proposition \ref{halfpint}}\label{paper}

The outline of the argument is as follows.

First, we think of $s(t)$, the realized proportion of cheaters at time $t$, as a function $s(s_1(t),t,\lll)$ of $s_1(t)$ 
(the insiders' perceived level of cheating at time $t$), $t$ and $\lll$.\footnote{In principle, we should write $s_1(t,\lll)$, 
but omit the latter argument to conserve on clutter.  We need not include $s_0(t)$ as an argument of $s$, since (from \eqref{diffs1s0}) 
this can be inferred from $s_1$ and $t$.}

Second, we show that for fixed $s_1$ and $t$,  the smaller is $\lll$ the smaller is $s(s_1,t,\lll))$.%
This in turn ensures that, for a fixed $s_1$ and $t$, the smaller is $\lll$, the smaller is $ds_1/dt$.

Third, suppose that the path of $s_1(t)$ induced by $\lll$ converges to $s_g$, the good equilibrium.  Then,  for a smaller 
value $\lll'$, we get a path in which, at every time $t$, either the induced value of $s_1$ is smaller, or (if equal) the derivative $ds_1/dt$ 
is smaller.  Hence, the path induced by the smaller value $\lll'$ is always either below or being pushed below that induced by $\lll$, 
and so the $\lll'$ path also converges to 0. Hence, if the path of $s_1(t)$ induced by $\lll$ converges to $s_g$, then so does the path induced by any $\lll'<\lll$.  
A similar argument shows that if the path of $s_1(t)$ induced by $\lll$ converges to $s_b$, then so does the path induced by any $\lll'>\lll$.  This gives [\ref{halfpint}.1].

Finally, we show [\ref{halfpint}.2], that at most one value $\lll\in [0,1/2]$ induces convergence to $s_u$.

We begin with a preliminary result.

\begin{lemma}\label{singlelamstar}
	Consider two paths of insider perceptions, $s_1(t,\lambda_1)$ and $s_1(t,\lambda_2)$, with $\lambda_1>\lambda_2$. Suppose both paths converge to $s_u$. Then, for all $t$ large enough, it must be the case that $s_1(t,\lambda_1)<s_1(t,\lambda_2)$, 
\end{lemma}	

\proof
To simplify the notation, denote by $s_1^j$ the path of the insider perceptions corresponding to $\lambda_j$.   For a $t$ large enough, we know that the dynamics of $s_1^1(t)$ and $s_1^2(t)$ follow
$$\dot s_1^1(t)=\ddd \left\{\frac12-\frac12 f(s_1^1(t))-s_1^1(t)+\frac{\lll_1}{2}(f(s_1^1(t))-f(s_1^1(t)+c))\right\},$$ 
and
$$\dot s_1^2(t)=\ddd \left\{\frac12-\frac12 f(s_1^2(t))-s_1^2(t)+\frac{\lll_2}{2}(f(s_1^2(t))-f(s_1^2(t)+c))\right\},$$
where $c=e^{-\delta t}s_b$ is, for a given $t$, a constant which is common to both paths.

We want to show that, if $t$ is large enough, it cannot be that $s_1^1(t)\geq s_1^2(t)$. Suppose, by way of contradiction, that this is the case. We will show that this implies that $$\dot s_1^1(t)> \dot s_1^2(t).$$
This in turn implies that $s_1^1(t)$ and $s_1^2(t)$ would diverge from each other, and therefore they could not both converge to $s_u$.

If at some (large) $t$ it were the case that $s_1^1(t)= s_1^2(t)$, it would follow (since $\lambda_1>\lambda_2$ and $f$ is decreasing) that $\dot s_1^1(t)> \dot s_1^2(t)$. Starting from $t$, the path for $s_1^1$ would then immediately be above the path for $s_1^2$. We would then need to consider the case $s_1^1(t)> s_1^2(t)$, to which we turn.

We have
\begin{eqnarray}\label{diffdots}
	\dot s_1^1(t)-\dot s_1^2(t)=\ddd \left\{\frac12 (f(s_1^2(t))-f(s_1^1(t)))+s_1^2(t)-s_1^1(t)+\right.\\
	\left.\frac{\lll_1}{2}(f(s_1^1(t))-f(s_1^1(t)+c))-\frac{\lll_2}{2}(f(s_1^2(t))-f(s_1^2(t)+c))\right\}\nonumber.
\end{eqnarray}
The second line tends to 0 as $t \to \infty$ but can be negative for a given $t$.  As a preliminary step, we show that
$$\frac12 (f(s_1^2(t))-f(s_1^1(t)))>s_1^1(t)-s_1^2(t).$$
Indeed,
\begin{equation}\nonumber
	\frac12 (f(s_1^2(t))-f(s_1^1(t))) = \frac{\qqq q}{2}\frac{(1-q)(s_1^1(t)-s_1^2(t))}{(q+(1-q)s_1^1(t))(q+(1-q)s_1^2(t))},
\end{equation}
hence $$\frac12 (f(s_1^2(t))-f(s_1^1(t)))>s_1^1(t)-s_1^2(t)$$ if $$\frac{\qqq q}{2}\frac{(1-q)}{(q+(1-q)s_1^1(t))(q+(1-q)s_1^2(t))}>1.$$
In turn, given that both  $s_1^1(t)$ and $s_1^2(t)$ are smaller than $s_u$, we have that
\begin{eqnarray}\nonumber
\begin{array}{lcr}
	\ds\frac{\qqq q}{2}\ds\frac{(1-q)}{(q+(1-q)s_1^1(t))(q+(1-q)s_1^2(t))}&>&\ds\frac{\qqq q}{2}\frac{(1-q)}{(q+(1-q)s_u)^2}\\
	& & \\
	&=&\ds\frac{\qqq q (1-q)}{2}\ds\frac{(1-2 s_u)^2}{(\qqq q)^2}\\
	& & \\
	&=&\ds\frac{(1-q)}{2\qqq q}(1-2 s_u)^2,
\end{array}
\end{eqnarray}
where we used equations (\ref{bandit}) and (\ref{brahms}) to replace $q+(1-q)s_u$.

Using now the definition of $s_u$ (equation (\ref{s_u})) we have that $$1-2 s_u=\frac{1+q+\sqrt{(q+1)^2-8\theta q(1-q)}}{2(1-q)}.$$
Therefore,
\begin{eqnarray}\nonumber
\begin{array}{rcl}
	\ds\frac{(1-q)}{2\qqq q}(1-2 s_u)^2&=&\ds\frac{2(1+q)^2-8 \qqq q(1-q)+2(1+q)\sqrt{(1+q)^2-8\qqq q(1-q)}}{8 \qqq q(1-q)}\\
	& &\\
	&=&\ds\frac{(1+q)^2}{4\qqq q(1-q)}-1+\frac{(1+q)}{4\qqq q(1-q)}\sqrt{(1+q)^2-8\qqq q(1-q)}.
\end{array}
\end{eqnarray}
We need to establish whether the right side is larger than 1. This is equivalent to establish whether
$$\sqrt{(1+q)^2-8\qqq q(1-q)}>\frac{8\qqq q(1-q)}{1+q}-(1+q).$$
Squaring both sides we obtain
$$(1+q)^2-8\qqq q(1-q)>\frac{(8\qqq q(1-q))^2}{(1+q)^2}+(1+q)^2-16\qqq q(1-q).$$
Simplifying this boils down to
$$(1+q)^2>8\qqq q(1-q),$$
which is a condition satisfied as long as we have 3 equilibria of the dynamic system. This establishes the preliminary step
\begin{eqnarray}\label{prelstep}
	\frac12 (f(s_1^2(t))-f(s_1^1(t)))>s_1^1(t)-s_1^2(t).
\end{eqnarray}

Rewrite now equation (\ref{diffdots}) as follows:
\begin{eqnarray}\label{diffdots1}
\begin{array}{l}
	\dot s_1^1(t)-\dot s_1^2(t)=\\ 
	\\
	\ddd \left\{\ds\frac12\{ [f(s_1^2(t))(1-\lll_2)+f(s_1^2(t)+c)\lll_2]
 -[f(s_1^1(t))(1-\lll_1)+f(s_1^1(t)+c)\lll_1]\}+s_1^2(t)-s_1^1(t)\right\}
\end{array}.	
\end{eqnarray}
The expression within the first pair of square brackets can be written as
\begin{eqnarray}\label{f2l}
	f(s_1^2(t))-\lll_2 k_2,
\end{eqnarray}
where
$$k_2=\frac{\qqq q(1-q)c}{(q+(1-q)s_1^2(t))(q+(1-q)(s_1^2(t)+c))}.$$
Similarly, the expression within the second pair of square brackets can be written as
\begin{eqnarray}\label{f1l}
	f(s_1^1(t))-\lll_1 k_1,
\end{eqnarray}
where
$$k_1=\frac{\qqq q(1-q)c}{(q+(1-q)s_1^1(t))(q+(1-q)(s_1^1(t)+c))},$$
and $k_2 > k_1$.

Therefore, the right side of (\ref{diffdots1}) can be written as
\begin{eqnarray}\nonumber
\ddd \left\{\frac12\{ f(s_1^2(t))-f(s_1^1(t))+\lll_1 k_1 -\lll_2 k_2\}+s_1^2(t)-s_1^1(t)\right\}.
\end{eqnarray}
We now show that, when $t$ is sufficiently large, and therefore $c$ is sufficiently small, $\lll_1 k_1 -\lll_2 k_2\geq 0$.
This inequality is equivalent to
\begin{eqnarray}\nonumber
\frac{\lll_1-\lll_2}{\lll_1}\geq\frac{k_2-k_1}{k_2}
=1-
\frac{(q+(1-q)s_1^2(t))(q+(1-q)(s_1^2(t)+c))}{(q+(1-q)s_1^1(t))(q+(1-q)(s_1^1(t)+c))}.
\end{eqnarray}

The left side is a positive, constant scalar. As $t$ becomes large the right side approaches 0. For a sufficiently large $t$ 
this then proves that $\lll_1 k_1 -\lll_2 k_2\geq 0$, which in turn implies, using
(\ref{prelstep}), that $\dot s_1^1(t)-\dot s_1^2(t)>0$.
\hfill\endproof

As we anticipated the actual proof of Proposition \ref{halfpint} is divided into four steps. 

\begin{step}{\rm
Recalling (\ref{dyn_s}) and using (\ref{diffs1s0}) (specialized to the case we are considering) we define
\begin{eqnarray}\label{eqn app: s in terms of zetas}
s(t)\coloneqq s(s_1(t),t,\lambda) = (1-\lll)^2\zeta_{11}+\lll(1-\lll)(\zeta_{01}+\zeta_{10})+\lll^2\zeta_{00}
\end{eqnarray}
where
\begin{eqnarray}\label{eqn app: zeta11}	
	\hspace{-1in}\zeta_{11}	= \min\left\{1,\max\{0,\max\{f(s_1(t)), \frac12+\frac12f(s_1(t))\}
	-f(s_1(t))\}\right\}
\end{eqnarray}
\begin{eqnarray}\label{eqn app: zeta10}
	\hspace{-1in}\zeta_{10}	 = \min\left\{1,\max\{0,\max\{f(s_1(t)), \frac12+\frac12f(s_1(t))\}
	-f(s_1(t)+e^{-\delta t}s_b)\}\right\}
\end{eqnarray}
\begin{eqnarray}\label{eqn app: zeta01}
	\zeta_{01}	 = \min\left\{1,\max\{0,\max\{f(s_1(t)+e^{-\delta t}s_b), \frac12+\frac12f(s_1(t)+e^{-\delta t}s_b)\}\right.
	f(s_1(t))\}\bigg\}
\end{eqnarray}
\begin{eqnarray}\label{eqn app: zeta00}
	\zeta_{00}	 = \min\left\{1,\max\{0,\max\{f(s_1(t)+e^{-\delta t}s_b), \frac12+\frac12f(s_1(t)+e^{-\delta t}s_b)\}\right.
	f(s_1(t)+e^{-\delta t}s_b)\}\bigg\}.
\end{eqnarray}
For any given $s_1(t)$ and $t$, we have
\begin{eqnarray}\nonumber
\begin{array}{lcr}
	\ds\frac{\partial s}{\partial \lll} & = & -2(1-\lll)\zeta_{11}+(1-2\lll)(\zeta_{01}+\zeta_{10})+2\lll \zeta_{00}\\
	&=& -2\zeta_{11}+(\zeta_{01}+\zeta_{10})+2\lll[\zeta_{11}+\zeta_{00}-(\zeta_{01}+\zeta_{10})]\\
	&&\\
	\ds\frac{\partial^2(s)}{\partial^2\lll}& = &2(\zeta_{11}+\zeta_{00}-(\zeta_{01}+\zeta_{10})).
\end{array}
\end{eqnarray}	
}
\end{step}

\begin{step}{\rm  We show that $\partial s/\partial \lll\ge0$ in the interval $\lll\in [0,1/2]$.
Because the second derivative has a constant sign over this interval, it suffices to show that $\partial s/\partial \lll\geq 0$ for $\lll=0$ and $\lll=\frac12$.
The corresponding requirements are
\begin{eqnarray}\label{eqn app: actual conditions for step 2}
\begin{array}{lcl}
	2\,\zeta_{11}&\leq &\zeta_{01}+\zeta_{10}\\
	\zeta_{11}&\leq&\zeta_{00}.
\end{array}	
\end{eqnarray}	
The second of these is almost immediate.\footnote{Intuitively, $\zeta_{11}$ is the level of cheating when two good agents meet, 
and $\zeta_{00}$ is the level of cheating when two bad agents meet.  The second requirement is then the statement that 
bad agents cheat more than good agents.} For any fixed $\lll$, for all $t$ it is the case that $s_1(t)+e^{-\delta t}s_b\geq s_1(t)$ (in fact the 
inequality is always strict and tends to an equality as $t$ tends to $\infty$). If $s_1(t)>s^*$, then also $s_1(t)+e^{-\delta t}s_b>s^*$. Therefore,
\begin{eqnarray}\nonumber
\zeta_{11}=\frac12-\frac12 f(s_1(t))<\zeta_{00}=\frac12-\frac12 f(s_1(t)+e^{-\delta t}s_b),
\end{eqnarray}
since $f$ is decreasing and $f(s_1(t))<1$. If $s_1(t)\leq s^*$, there are two possibilities: either
$s_1(t)+e^{-\delta t}s_b>s^*$ or $s_1(t)+e^{-\delta t}s_b\leq s^*$. In the first case, %
\begin{eqnarray}\nonumber
\zeta_{11}=0<\frac12-\frac12 f(s_1(t)+e^{-\delta t}s_b)=\zeta_{00}.
\end{eqnarray}
 In the second case,
 \begin{eqnarray}\nonumber
 \zeta_{11}=0=\zeta_{00}.
 \end{eqnarray}

Moving to the first, we need $2 \,\zeta_{11}\leq \zeta_{01}+\zeta_{10}$. We can simplify the expressions for $\zeta_{11}$, $\zeta_{01}$ and $\zeta_{10}$ as follows
(for notational convenience, we neglect the dependence of $s_1$ on $t$ and we denote by $s_0$ the term $s_1(t)+e^{-\delta t}s_b$):
		\begin{eqnarray}\label{eqn app: thee zetas}
	         \begin{array}{rcl}		
			\zeta_{11}	&=& \max\left\{0,\ds\frac12-\ds\frac12f(s_1)\right\}\\
			
			&&\\
			\zeta_{10}	 &=& \min\left\{1,\max\{f(s_1)-f(s_0), \ds\frac12+\ds\frac12f(s_1)-f(s_0)\}\right\}\\
			
			&&\\
			\zeta_{01}	 &=& \max\left\{0, \ds\frac12+\ds\frac12f(s_0)-f(s_1)\right\},
		\end{array}	
		\end{eqnarray}
		These hold  because,
		
		\begin{itemize}
			\item In equation (\ref{eqn app: zeta11}) for $\zeta_{11}$, if $s_1\leq s^*$, the inner maximum is solved by $f(s_1)$, so the whole expression is 0, 
                          while if $s_1> s^*$ the inner maximum is solved by $\frac12 +\frac12 f(s_1)<1$, so the whole expression is $\frac12-\frac12f(s_1)$;
			
			\item In equation (\ref{eqn app: zeta10}) for 
			$\zeta_{10}$, again, if $s_1\leq s^*$, the inner maximum is solved by $f(s_1)$, hence we have $f(s_1)-f(s_0)$; 
			this could be bigger than $1$, so we cannot neglect the outer minimum. If $s_1> s^*$ the inner maximum is solved by $\frac12 +\frac12 f(s_1)$, 
			so the whole expression is $\frac12+\frac12f(s_1)-f(s_0)$; since $1>f(s_1)>f(s_0)$, this is positive;
			
			\item In equation (\ref{eqn app: zeta01}) for $\zeta_{01}$, if $s_0\leq s^*$, the inner maximum is solved by $f(s_0)$, hence we have $f(s_0)-f(s_1)$; this is negative, 
			so we need to bound the whole expression below by zero. If $s_0> s^*$ the inner maximum is solved by $\frac12 +\frac12 f(s_0)$. 
			We then have $\frac12+\frac12f(s_0)-f(s_1)$, which also could be negative, since $f(s_1)$ could be bigger than 1 (if $s_1<s^*$) and anyway is bigger than $f(s_0)$.
		\end{itemize}
	
		The expression
		$2 \,\zeta_{11}\le \zeta_{01}+\zeta_{10}$ can now be written as
		\begin{eqnarray}\label{eqn app: it all boils down plus inequality}
		\begin{array}{rcc}
			\max\{0,1-f(s_1)\}&\le&\min\left\{1,\max\{f(s_1)-f(s_0), \ds\frac12+\frac12f(s_1)-f(s_0)\}\right\}\\
			&&\\
			&&~~~+~\max\left\{0,\ds\frac12+\ds\frac12f(s_0)-f(s_1)\right\}
		\end{array}.
		\end{eqnarray}
		If the maximum on the left side of  (\ref{eqn app: it all boils down plus inequality}) 
		is zero, the inequality is satisfied and we have that both conditions in (\ref{eqn app: actual conditions for step 2}) are true. Let us then assume that the second maximum on the left side is positive.  
		This is equivalent to $f(s_1)<1$, and so we now maintain this assumption.  This in turn ensures that the minimum in the first term on 
		the right side of  (\ref{eqn app: it all boils down plus inequality}) is not 1 and the first maximum is realized by its second term, and so we have
		\[
		1-f(s_1)\le\left [\ds\frac12+\ds\frac12f(s_1)-f(s_0)\right]+\max\left\{0, \ds\frac12+\ds\frac12f(s_0)-f(s_1)\right\}.
		\]
		To prove the second condition in (\ref{eqn app: actual conditions for step 2}) it then suffices to show that this inequality holds no matter which term in the final maximum is larger, which is equivalent to
		\begin{eqnarray}\label{eqn app: buffer}
		1-f(s_1) \le \left\{
		\begin{array}{l}
			\ds\frac12+\ds\frac12f(s_1)-f(s_0)\\
			\\
			1-\ds\frac12f(s_1)-\ds\frac12f(s_0).	
		\end{array}	
		\right.
		\end{eqnarray}
		The second of these simplifies to $0\le (f(s_1)-f(s_0))$, which is always true.  We thus need to check the first, which is
		\begin{eqnarray}\nonumber
		\frac12\le \frac32f(s_1)-f(s_0),
		\end{eqnarray}
		or, equivalently,
		\begin{eqnarray}\nonumber
		\frac12(1-f(s_1))\le f(s_1)-f(s_0).
		\end{eqnarray}
		Remember, however, that we are considering the case  when 0 is larger than $\frac12+\frac12f(s_0)-f(s_1)$, and hence $f(s_0)< 2f(s_1)-1$, 
		which is equivalent to $f(s_1)-f(s_0) >1-f(s_1)$. Since we are considering the case $f(s_1)<1$, we then have
		\begin{eqnarray}\nonumber
		\frac12(1-f(s_1))<1-f(s_1)< f(s_1)-f(s_0),
		\end{eqnarray}
		which is the first in (\ref{eqn app: buffer}). Therefore both conditions in (\ref{eqn app: actual conditions for step 2}) are satisfied. 
		
		Hence, in the interval $\lll \in [0,\frac12]$, for a fixed $s_1$ and $t$, we have ${\partial s}/{\partial \lambda} \geq 0$.
		This in turn ensures, given that $\dot{s_1}(t)$ is increasing in $s$, that for a fixed $s_1$ and $t$, the smaller is $\lll$, 
		the smaller is $\dot{s_1}(t)$.
	}	
		\end{step}
		
		\begin{step}{\rm
		Now consider a $\lll\leq \frac12$ such that the path of $s_1(t)$ converges to 0, the good equilibrium
		and take a smaller value $\lll'$.
		
		At time 0 and initial condition $s_1(0)=0$, common for both $\lll$s, we now know that $s(0,\lll)>s(0,\lll')$.
		Hence the path of $s_1$ induced by $\lll'$ is initially below the path induced by $\lll$.
		
		If the former path always remained weakly below the latter, it would also converge to 0.
		
		By contradiction, suppose it does not converge to 0. Then there must be a (finite) $t$ such that the path induced by $\lll'$ crosses, from below, the path induced by $\lll$. At that $t$, $s_1(t,\lll')=s_1(t,\lll)$. Hence, given $t$ and this value for $s_1$,
		we have that
		 $$\frac{d s_1(t,\lll')}{d t}\le\frac{d s_1(t,\lll)}{d t}.$$
		Hence, the path induced by the smaller value  $\lll'$ is always either below or being pushed below that induced by $\lll$, and so the path
		induced by $\lll'$ also converges to 0.  A similar argument shows that if the path of $s_1(t)$ induced by $\lll$ 
		converges to $s_b$, then so does the path induced by any $\lll>\lll'$.
}		
\end{step}		

\begin{step}{\rm
Now consider [\ref{halfpint}.2].  Suppose we have two paths, $s_1(t,\lambda_1)$ and $s_1(t,\lambda_2)$, with $\frac{1}{2}\geq \lambda_1>\lambda_2$, both 
converging to $s_u$.  Our previous steps show that the first path (associated to the larger $\lambda$) must always lie at least weakly  
above the second path (associated to the smaller $\lambda$).  
Using Lemma \ref{singlelamstar}  we then have a contradiction and hence the proof is now complete.\hfill\endproof	
}
\end{step}

\section{Proof of Corollary \ref{corollary: swap insiders and outsiders}}\label{section app: proof of corollary}

Proposition \ref{dvorak} established that for any $\lll >0$ there exists a $q^*$ $>$ $0$ such that, for any $q \leq q^*$, 
the system converges to $s_b$. Defining $\hat{\lll}=1-\lll$, this also means that for any $\hat{\lll}<1$ there exists a $q^*$ 
such that, for any $q \leq q^*$, the system converges to $s_b$. This is the claim we wanted to establish.
\hfill\endproof

\end{appendix}

\end{document}